\documentclass{aa}
\usepackage[varg]{txfonts}
\usepackage[separate-uncertainty=true,
            multi-part-units=single]{siunitx}
\usepackage[version=3]{mhchem}
\usepackage{amsmath}
\usepackage{lscape}
\usepackage{graphicx}

\def\aap{A\&A}

\def\apjs{ApJS}
\def\apjl{ApJL}

\bibpunct{(}{)}{;}{a}{}{,}

\begin{document}

\title{SWEET-Cat update and FASMA}
\subtitle{A new minimization procedure for stellar parameters using high-quality spectra\thanks{Based on
observations collected at the La Silla Observatory, ESO (Chile), with FEROS/2.2m
(run 2014B/020), with UVES/VLT at the Cerro Paranal Observatory (runs ID
092.C-0695, 093.C-0219, 094.C-0367, 095.C-0324, and 096.C-0092), and with
FIES/NOT at Roque de los Muchachos (Spain) (runs ID 14AF14 and 53-202).},\thanks{The compiled SWEET-Cat is available online.}}

\author{ D.~T.~Andreasen\inst{1,2}
    \and S.~G.~Sousa\inst{1}
    \and M.~Tsantaki\inst{3}
    \and G.D.C.~Teixeira\inst{1,2}
    \and A.~Mortier\inst{4}
    \and N.~C.~Santos\inst{1,2}
    \and L.~Su\'arez-Andr\'es\inst{5,6}
    \and E.~Delgado-Mena\inst{1}
    \and A.~C.~S.~Ferreira\inst{1,2}
}

\institute{
Instituto de Astrof\'isica e Ci\^encias do Espa\c{c}o, Universidade do Porto, CAUP, Rua das Estrelas, 4150-762 Porto, Portugal
\email{daniel.andreasen@astro.up.pt}
\and
Departamento de F\'isica e Astronomia, Faculdade de Ci\^encias, Universidade do Porto, Rua Campo Alegre, 4169-007 Porto, Portugal
\and
Instituto de Radioastronom\'ia y Astrof\'isica, IRyA, UNAM, Campus Morelia, A.P. 3-72, 58089 Michoac\'an, Mexico
\and
Centre for Exoplanet Science, SUPA, School of Physics and Astronomy, University of St Andrews, St Andrews KY16 9SS, UKSUPA, School of Physics and Astronomy, University of St Andrews, St Andrews KY16 9SS, UK
\and
Depto. Astrof\'isica, Universidad de La Laguna (ULL), E-38206 La Laguna, Tenerife, Spain
\and
Instituto de Astrof\'isica de Canarias, E-38205 La Laguna, Tenerife, Spain
}

\date{Received ...; accepted ...}

\abstract
{Thanks to the importance that the star-planet relation has to our understanding
of the planet formation process, the precise determination of stellar parameters
for the ever increasing number of discovered extrasolar planets is of great
relevance. Furthermore, precise stellar parameters are needed to fully
characterize the planet properties. It is thus important to continue the efforts
to determine, in the most uniform way possible, the parameters for stars with
planets as new discoveries are announced.}
{In this paper we present new precise atmospheric parameters for a sample of 50
stars with planets. The results are presented in the catalogue: SWEET-Cat.}
{Stellar atmospheric parameters and masses for the 50 stars were derived
assuming local thermodynamic equilibrium (LTE) and using high-resolution and
high signal-to-noise spectra. The methodology used is based on the measurement
of equivalent widths with ARES2 for a list of iron lines. The line
abundances were derived using MOOG. We then used the curve of growth analysis to
determine the parameters. We implemented a new minimization procedure which
significantly improves the computational time.}
{The stellar parameters for the 50 stars are presented and compared with
previously determined literature values. For SWEET-Cat, we compile values
for the effective temperature, surface gravity, metallicity, and stellar mass
for almost all the planet host stars listed in the Extrasolar Planets
Encyclopaedia. This data will be updated on a continuous basis. The data can be
used for statistical studies of the star-planet correlation, and for the
derivation of consistent properties for known planets.}
{}

\maketitle

\section{Introduction}
\label{sec:introduction}
The study of extrasolar planetary systems is an established field of research.
To date, more than 3500 extrasolar planets have been discovered around more than
2500 solar-type stars\footnote{For an updated table we refer to
\url{http://www.exoplanet.eu}}. Most of these planets have been found thanks to the
incredible precision achieved in photometric transit and radial velocity
methods. Not only do we have intriguing new types of planetary systems that
challenge current theories, but the increasing number of exoplanets also allows us
to do statistical studies of the new-found worlds by analysing their atmospheric
composition, internal structure, and planetary composition.

Precise and accurate planetary parameters (mass, radius, and mean density) are
needed to distinguish between solid rocky, water rich, gaseous, or otherwise
composed planets. A key aspect to this progress is the characterization of the
planet host stars. For instance, precise and accurate stellar radii are critical
if we want to measure precise and accurate values of the radius of a transiting
planet \citep[see e.g.][]{Torres2012,Mortier2013}. The determination of the
stellar radius is in turn dependent on the quality of the derived stellar
atmospheric parameters such as the effective temperature.

We continue the work of \citet{Santos13} by deriving atmospheric parameters,
namely the effective temperature ($T_\mathrm{eff}$), surface gravity ($\log g$),
metallicity ([Fe/H], where iron often is used as a proxy for the total
metallicity), and the micro turbulence ($\xi_\mathrm{micro}$) for a sample of
planet host stars. This, in turn, allows us to study new correlations between
planets and their hosts in a homogeneous way or to gain higher statistical
certainty on the already discovered correlations.

The analysis of high-quality spectra, i.e. spectra with high spectral resolution
and a high signal-to-noise ratio (S/N), plays an important role in the
derivation of stellar atmospheric parameters. Nevertheless, spectral analysis is
a time consuming method. There has been an increase in the number of optical
high-resolution spectrographs available and, additionally, a number of near-IR
spectrographs are either planned or are already available making the task of
analysing the increasing amount of spectra even more crucial.

In the era of large data sets, computation time has to be decreased as much as
possible without compromising the quality of the results. In the light of this
we have developed a tool, FASMA, for deriving atmospheric parameters in a fast
and robust way using standard spectroscopic methods. We made this tool available
as a web interface\footnote{\url{http://www.iastro.pt/fasma}}. This works well
for optical spectra, which we demonstrate in Section~\ref{sub:Testing_FASMA}
using the line list from \citet{Sousa2008a} and \citet{Tsantaki2013}. This tool
also ships with a line list for near-IR spectra using the line list presented
recently in \citet{Andreasen2016}. The tool is provided to the community as an
easy to use web tool to avoid any problems with installations. The tool is
described in detail in Section~\ref{sec:FASMA}. In Section~\ref{sec:results} we
present the new parameters for SWEET-Cat.

\section{FASMA}
\label{sec:FASMA}
Fast Analysis of Spectra Made Automatically (FASMA) is a web
tool\footnote{\url{http://www.iastro.pt/fasma}} for analysing spectra. FASMA is
written in Python and works as a wrapper around ARES2 \citep{Sousa2015a}
(hereafter just ARES) and MOOG \citep[][version 2014]{Sneden1973}, for an
all-in-one tool. ARES is a tool used to automatically measure equivalent widths (EW)
from a spectrum given a line list. MOOG is a radiative transfer code under the
assumption of local thermodynamic equilibrium (LTE).

FASMA has three different drivers: i) measuring EWs using ARES, ii) deriving
stellar parameters from a set of measured \ion{Fe}{I} and \ion{Fe}{II} line EWs
(tested extensively on FGK dwarf and subgiant stars), and iii) deriving
abundances for 15 elements, all described below. The model atmospheres are
formatted in a grid of Kurucz Atlas 9 plane-parallel, 1D static model
atmospheres \citep{Kurucz1993}. FASMA can also manage the new grid of Atlas
models calculated by \citet{Meszaros2012} for the APOGEE survey and the MARCS
models \citep{Gustafson2008}. The interpolation from the grid is calculated from
a geometric mean for effective temperature, surface gravity, and metallicity.

We do not consider hyper-fine structure (HFS) when deriving abundances since it
has little or no effect on the derivation of iron abundances, so the derived
parameters are trustworthy. If necessary, we might implement this in the future
for elements where HFS is important.

\subsection{Equivalent width measurements}
\label{sub:EW_measurements}
The EWs are strongly correlated with the atmospheric parameters. Measurements of
the EW can be done manually using a tool like splot in IRAF, but often when
dealing with a large sample of stars this is not a suitable way to deal with the
task. Therefore, there are several tools like ARES which can measure the EWs of
spectral lines automatically. To use this mode of FASMA, a line
list\footnote{ARES, in principle, just needs a list of wavelengths in order to
run, but is often used with a line list with characteristics of the atomic
absorption line.} and a spectrum (the format should be 1D fits for ARES to read
it) are needed. FASMA is shipped with some line lists ready to use. The output
will be a line list with the newly measured EWs in the format required for MOOG.
The output can be used for either stellar parameter derivation or the abundance
method, both described below. ARES iterates over all the lines to be measured.
For each line a small window is selected, where a local normalization of the
spectrum is made automatically. The normalization is made based on a range of
settings described in \citet{Sousa2007,Sousa2015a}. All of these settings can be
changed in the driver. Most important is the \emph{rejt} parameter, which we
recommend measuring using the S/N of the spectrum. We note that for late K
dwarfs and cooler, ARES starts to have difficulties placing the continuum. If
the user wants EWs that are as accurate as possible, we recommend measuring by
hand for these stars. The \emph{rejt} parameter is a number between 0 and 1.
When the value is closer to 1, which means higher quality of the spectrum, the
normalization will use the higher points in the spectrum. By using the S/N, the
\emph{rejt} is simply, $\mathit{rejt}=1-1/S/N$.

The line lists shipped with FASMA are presented in Table~\ref{tab:linelists}.
These line lists are all calibrated for the Sun, i.e. the oscillator strengths
for each absorption line are changed so the line with the measured EW from a
solar spectrum return solar abundance for the given element.

\begin{table}[htb!]
    \caption{Line lists provided with FASMA. The first three line lists
             are for parameter determination while the last line list is
             used to derive abundances for 15 different elements.}
    \label{tab:linelists}
    \centering
    \begin{tabular}{lrrl}
      \hline\hline
      Line list             & \ion{Fe}{I}/\ion{Fe}{II} & Elements   & Usage      \\
      \hline
      \citet{Sousa2008a}    &  263/36                  &  1         & Parameters \\
      \citet{Tsantaki2013}  &  120/17                  &  1         & Parameters \\
      \citet{Andreasen2016} &  249/5                   &  1         & Parameters \\
      \citet{Neves2009}     &  -/-                     & 15         & Abundances \\
      \hline
    \end{tabular}
\end{table}

\subsection{Stellar parameter derivation}
\label{sub:EW_method}
The standard determination of spectroscopic parameters for solar-type stars
starts by measuring the EW of selected and well-defined absorption lines. Then
we translate these measurements into individual line abundances, assuming a
given atmospheric model. We obtain the correct stellar parameters by imposing
excitation and ionization balance for the iron species. This is a classical
curve-of-growth analysis using the Boltzmann and Saha equations,
\begin{align}
  \frac{N_n}{N} &= \frac{g_n}{u(T)}10^{-\theta \chi_n} \tag*{Boltzmann} \\
  \frac{N_1}{N_0}P_e &= \frac{(2\pi m_e)^{3/2}(kT)^{5/2}}{h^3} \frac{2u_1(T)}{u_0(T)} e^{-I/kT}, \tag*{Saha}
\end{align}
where $N$ is the number of particles per unit volume, $N_n$ is the fraction of
atoms/ions excited to the $n$th state, $g_n$ is the statistical weight,
$\theta=5040/T$, $T$ is the temperature, $u(T)$ are the so-called partition
functions, $m_e$ is the electron mass, $P_e$ is the electron pressure, and $I$
is the ionization potential. More details can be found in e.g. \citet{Gray2006}.
We note that the abundance determination are calculated through the MOOG code:

\begin{itemize}
    \item The effective temperature has a strong influence on the correlation
          of iron abundance with the excitation potential (excitation balance).
          We obtain the $T_\mathrm{eff}$ when \ion{Fe}{I} abundance shows no
          dependence on the excitation potential, i.e. the slope of abundance
          versus excitation potential is zero.
    \item Surface gravity is derived from the ionization balance of \ion{Fe}{I}
          and \ion{Fe}{II} abundances. Therefore, the abundance of neutral iron
          should be equal to the abundance of ionized iron and consistent with
          that of the input model atmosphere.
    \item Microturbulence is connected with the saturation of the stronger iron
          lines. However, the abundances for weak and strong lines of a certain
          species (in our case iron) should be the same independent of the value
          of $\xi_\mathrm{micro}$. Iron abundances should show no dependence on
          the reduced EW ($\log(EW/\lambda)$), i.e. the slope of abundance
          versus the reduced EW is zero.
\end{itemize}
Lastly, we change the input [\ion{Fe}/\ion{H}] to match that of the average
output [\ion{Fe}/\ion{H}]. Hence we have four criteria to minimize
simultaneously:

\begin{enumerate}
    \item The slope between abundance and excitation potential ($a_\mathrm{EP}$)
          has to be lower than 0.001.
    \item The slope between abundance and reduced EW ($a_\mathrm{RW}$) has to be
          lower than 0.003. We use 0.003 rather than 0.001 since this slope
          varies more rapidly with small changes in atmospheric parameters.
    \item The difference between the average abundances of \ion{Fe}{I} and
          \ion{Fe}{II} ($\Delta\ion{Fe}{}$) should be less than 0.01.
    \item Input and output metallicity should be equal.
\end{enumerate}
These criteria are used as indicators for the physical parameters we are trying
to minimize. The constraints on the first three indicators are empirically
defined.

\begin{figure}[tpb]
    \centering
    \includegraphics[width=1.0\linewidth,natwidth=700,natheight=650]{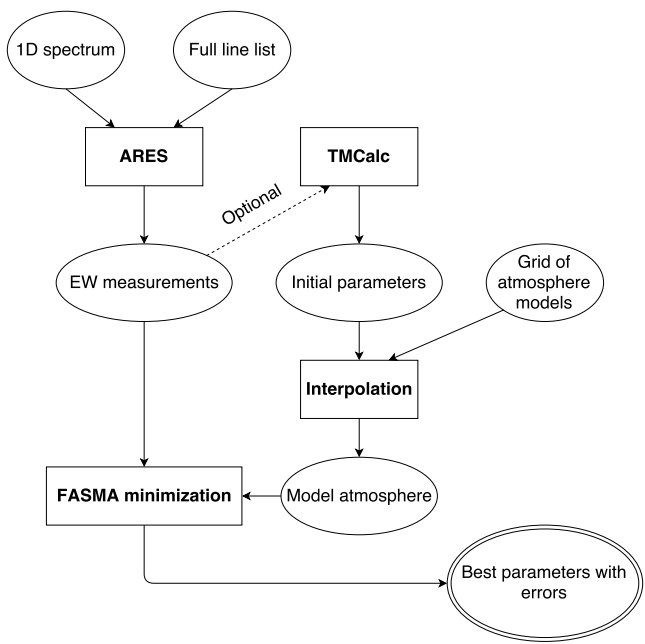}
    \caption{General overview of FASMA from spectrum to parameters.}
    \label{fig:FASMA_general}
\end{figure}

\begin{figure}[tpb]
    \centering
    \includegraphics[width=1.0\linewidth,natwidth=700,natheight=650]{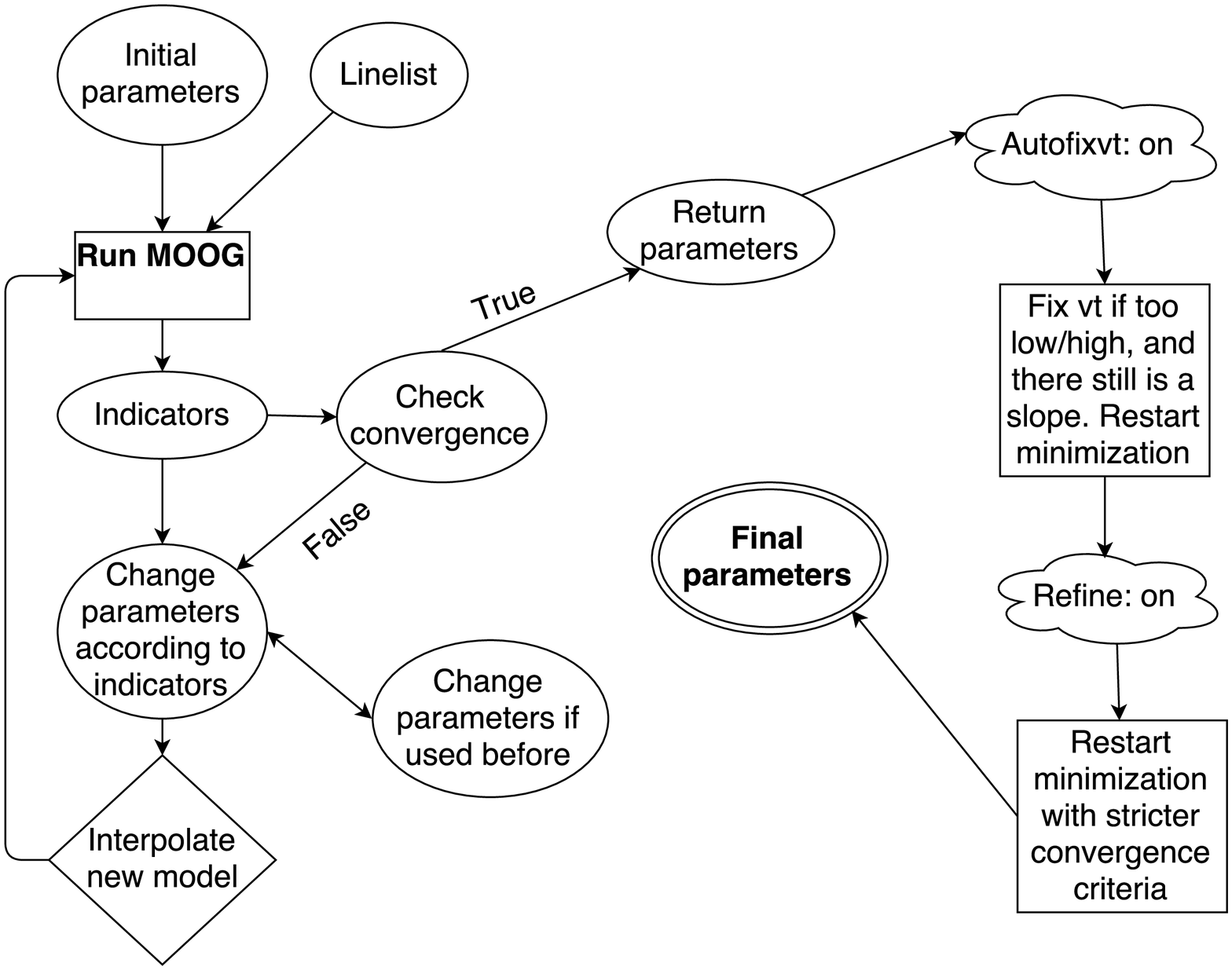}
    \caption{Overview of the minimization for FASMA with the
    EW method.}
    \label{fig:FASMA_minimization}
\end{figure}

There are many minimization routines available in Python. The ones from the
SciPy ecosystem\footnote{\url{http://scipy.org}} are the most commonly used.
There are some advantages and disadvantages to using proprietary minimization
routines. The advantages are that it is already written, and there is usually
good documentation for libraries such as SciPy. The disadvantage in this
situation is that most minimization routines do not work well with vector
functions returning another vector:
\begin{align}
    f(\{T_\mathrm{eff}, \log g, [Fe/H], \xi_\mathrm{micro}\}) = \{a_\mathrm{EP}, a_\mathrm{RW}, \Delta\ion{Fe}, \ion{Fe}{I}\}. \label{eq:vector}
\end{align}
A workaround to this is to combine the criteria into one single criterion, for
example by adding them quadratically and minimizing that expression instead.
Thus, we have a vector function returning a scalar:
\begin{align}
    f(\{T_\mathrm{eff}, \log g, [Fe/H], \xi_\mathrm{micro}\}) &= \sqrt{a_\mathrm{EP}^2 + a_\mathrm{RW}^2 + \Delta\ion{Fe}{}^2}. \label{eq:scalar}
\end{align}
The minimization routines are also not physical in the sense that they are not
written for the problem. These two disadvantages encouraged us to write a
specific minimization routine for the problem at hand. This also allowed us to
minimize the more complicated expression in Equation~\ref{eq:vector}. Here is
how it works:

\begin{enumerate}
    \item Run MOOG once with user defined initial parameters (default is
          solar) and calculate $a_\mathrm{EP}$, $a_\mathrm{RW}$, and
          $\Delta$\ion{Fe}.
    \item Change the atmospheric parameters ($T_\mathrm{eff}$, $\log g$,
          [\ion{Fe}/\ion{H}], $\xi_\mathrm{micro}$) according to the size of the
          indicator. A parameter is only changed if it is not fixed.
    \begin{itemize}
        \item $a_\mathrm{EP}$: Indicator for $T_\mathrm{eff}$. If this value
              is positive, then increase $T_\mathrm{eff}$. Decrease
              $T_\mathrm{eff}$ if $a_\mathrm{EP}$ is negative.
        \item $a_\mathrm{RW}$: Same as above, but for $\xi_\mathrm{micro}$.
        \item $\Delta$\ion{Fe}{}: Positive $\Delta$\ion{Fe}{} means $\log g$
              should be decreased and vice versa.
        \item $[\ion{Fe}{}/\ion{H}{}]$ is changed to the output
              $[\ion{Fe}{}/\ion{H}{}]$ in each iteration.
    \end{itemize}
    \item If the new set of parameters have already been used in a previous
          iteration, then change them slightly. This is done by drawing a random
          number from a Gaussian distribution with a mean at the current value
          and a sigma equal to the absolute value of the indicator. For
          $T_\mathrm{eff}$ the new value would be a random draw from
          $f(x|T_\mathrm{eff,old},a_\mathrm{EP}^2) = \frac{1}{\sqrt{2\pi a_\mathrm{EP}^2 }}e^{-\frac{(x - T_\mathrm{eff,old})^2 } {2 a_\mathrm{EP}^2} }$
          and similar for the other atmospheric parameters using the appropriate
          indicators.
    \item Calculate a new atmospheric model by interpolating a grid of models
          so we have the requested parameters and run MOOG once again.
    \item For each iteration save the parameters used and the quadratic sum of
          the indicators.
    \item Check for convergence, i.e. if the indicators are below or equal
          to the empirical constraints chosen. If we do not reach convergence,
          then return the best found parameters. The best found parameters,
          when convergence is not reached, are chosen when the quadratic sum
          of the indicators are smallest.
\end{enumerate}
This whole process is shown schematically in Figure~\ref{fig:FASMA_general}, and
the minimization routine itself in Figure~\ref{fig:FASMA_minimization}. By
minimizing Equation~\ref{eq:vector} rather than Equation~\ref{eq:scalar} we can
reach convergence more quickly since we know in which direction we must change
the atmospheric parameters. The stepping in parameters follows these simple
empirical equations where we add the right side (sign change according to the
sign of the indicator) to the left side:
\begin{align}
    T_\mathrm{eff}     &: \SI{2000}{K} \cdot a_\mathrm{EP}   \\
    \xi_\mathrm{micro} &: \SI{1.5}{km/s} \cdot a_\mathrm{RW} \\
    \log g             &: -\Delta\ion{Fe}.
\end{align}
The metallicity is corrected at each step so the input metallicity matches that
of the output metallicity of the previous iteration. The functional form
(linear) for changing the parameters were found by changing one parameter, e.g.
$T_\mathrm{eff}$, while keeping the other parameters fixed at their convergence
values using the Sun as an example. A linear fit was applied to $T_\mathrm{eff} -
T_\mathrm{eff,0}$ versus $a_\mathrm{EP}$ in order to get the slope (\SI{2000}{K}
for $T_\mathrm{eff}$). Since we ignore all interdependencies between the
parameters, we slightly lowered the slopes found and arrived at the very simple
equations above. By empirically determining how the atmospheric parameters
should be changed in each iteration, we are able to swiftly approach the
convergence value. The error estimates are based on the same method presented in
\citet{Gonzalez2000}, which is also described in detail in \citet{Santos2003}
and \citet{Andreasen2016}.

By using the indicators like this, we can reach convergence quickly. The typical
calculation time for an FGK dwarf with a high-quality spectrum is around
$\SI{2}{min}$.

It is possible to run the EW method with a set of different options which are
described here.
\begin{itemize}
    \item \emph{fixteff}: This option fixes $T_\mathrm{eff}$ and derives the
          other parameters. The same is available for $\log g$ (\emph{fixlogg}),
          [\ion{Fe}/\ion{H}] (\emph{fixfeh}), and $\xi_\mathrm{micro}$
          (\emph{fixvt}). One or more parameters can be fixed. When one or more
          parameters are fixed, the corresponding indicator will be ignored for
          each iteration, thus the parameter itself will not be changed.
    \item \emph{outlier}: Remove a spectral line (or lines) after the the
          minimization is done if the abundance of this spectral line is more
          than $3\sigma$ away from the average abundance of all the lines. After
          the removal of outlier(s) the minimization routine restarts. The
          options are to remove all outliers above $3\sigma$ once or
          iteratively, or to remove one outlier above $3\sigma$ once or iteratively.
    \item \emph{autofixvt}: If the minimization routine does not converge and
          $\xi_\mathrm{micro}$ is close to 0 or 10 with a significant
          $a_\mathrm{RW}$ (numerically bigger than 0.05), then fix
          $\xi_\mathrm{micro}$. This option was added since we saw this
          behaviour in some cases. The solution was typically to restart the
          minimization manually with $\xi_\mathrm{micro}$ fixed. If
          $\xi_\mathrm{micro}$ is fixed it is changed at each iteration
          according to an empirical relation. For dwarfs it follows the one
          presented in \citet{Tsantaki2013} and for giants it follows the one
          presented in \citet{Adibekyan2015}.
    \item \emph{refine}: After the minimization is done, run it again from the best
          found parameters but with stricter criteria. If this option is set,
          it will always be the last step (after removal of outliers). The
          convergence criteria can be changed by the user, but we recommend
          using the defaults provided above.
    \item \emph{tmcalc}: Use TMCalc \citep{Sousa2012} to quickly estimate the
          $T_\mathrm{eff}$ and $[\ion{Fe}/\ion{H}]$ using the raw output from
          ARES. We then assume solar surface gravity ($\SI{4.44}{dex}$) and
          estimate $\xi_\mathrm{micro}$ based on an empirical relation (see above).
\end{itemize}

For the optical we used the line list presented in \citet{Sousa2008a}. However,
this line list does not work well for cool stars. This was fixed in
\citet{Tsantaki2013} by removing some lines from \citet{Sousa2008a}. For stars
cooler than \SI{5200}{K} we automatically rederived the atmospheric parameters
after removing lines so the line list resembled that of \citet{Tsantaki2013}.
The line list for the near-IR is also available \citep{Andreasen2016}.

All restarts of the minimization routine are done using the most recently found
best parameters as initial conditions.

\subsection{Abundance method}
\label{sub:Abundance_method}

FASMA calculates element abundances for 12 different elements (\ion{Na}{},
\ion{Mg}{}, \ion{Al}{}, \ion{Si}{}, \ion{Ca}{}, \ion{Ti}{}, \ion{Cr}{},
\ion{Ni}{}, \ion{Co}{}, \ion{Sc}{}, \ion{Mn}{}, and \ion{V}{}) from spectral
lines determined in \citet{Neves2009} and \citet{Adibekyan2012}. It also
includes three ionized elements: \ion{Cr}{II}, \ion{Sc}{2}, and \ion{Ti}{II}.
The abundances are derived using MOOG in the same way as described above for
\ion{Fe}{}. The atomic data were calibrated with the Sun as reference and solar
abundances from \citet{Anders1989}. The EWs are measured automatically with the
ARES driver of FASMA. The element abundance of each line is derived using the
atmospheric parameters of the stars obtained from the previous step. The final
element abundance is calculated from the weighted mean of the abundances
produced by all lines detected for a given element as described in
\citet{Adibekyan2015b}.

\subsection{Testing FASMA}
\label{sub:Testing_FASMA}

\begin{figure*}[tpb]
    \centering
    \includegraphics[width=1.0\linewidth,natwidth=750,natheight=500]{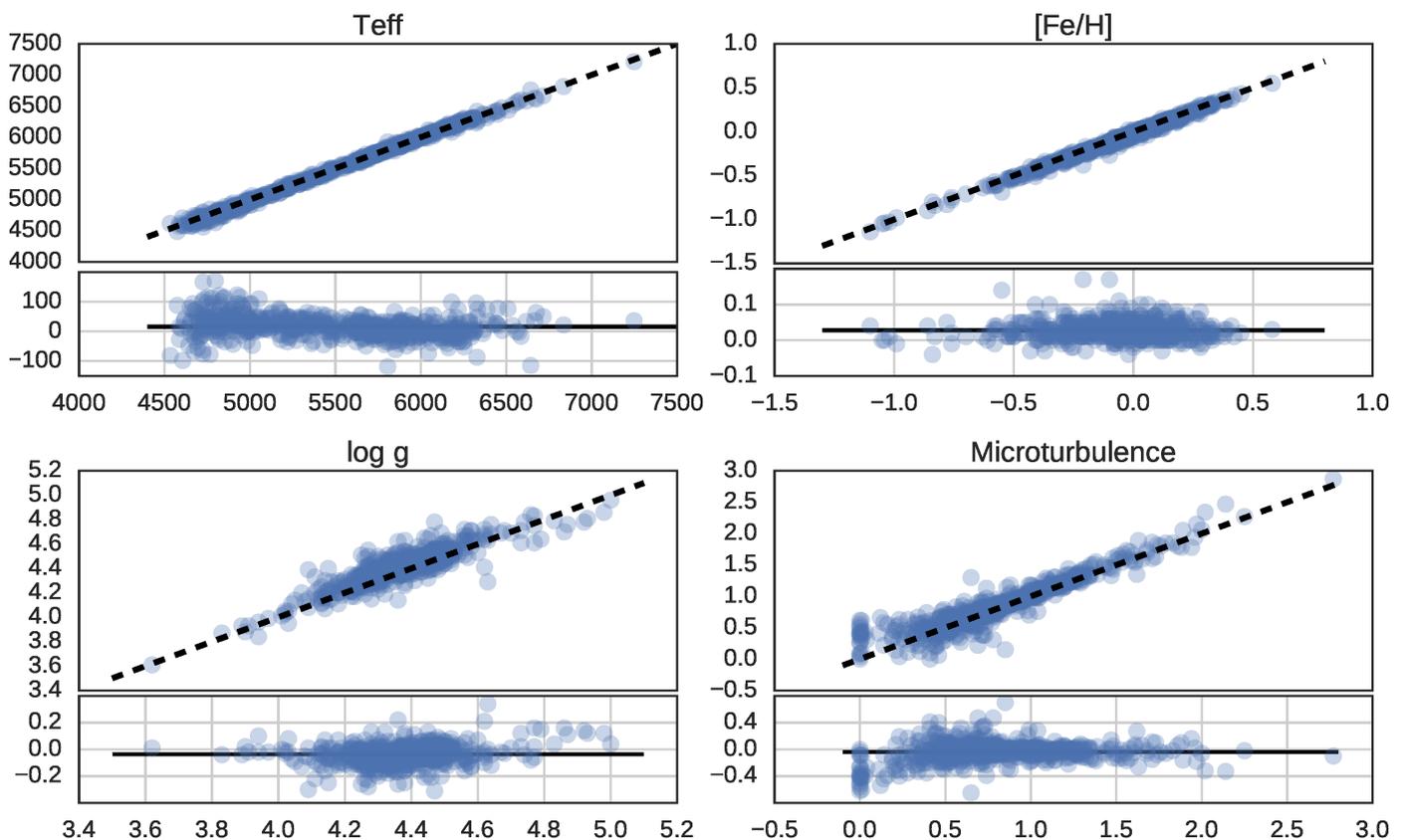}
    \caption{Stellar atmospheric parameters derived by FASMA compared
    to the sample by \citet{Sousa2011}. The x-axis in all plots shows the results
    from FASMA, while the y-axis shows the parameters derived by \citet{Sousa2011}.}
    \label{fig:FASMATest}
\end{figure*}

To test the derivation of stellar parameters implemented in FASMA we derived
parameters from the 582 sample by \citet{Sousa2011}. We used ARES to measure the
EWs. ARES can give an estimate of the S/N by analysing the continuum in certain
intervals. For solar-type stars the following intervals worked well:
\SIrange{5764}{5766}{\angstrom}, \SIrange{6047}{6053}{\angstrom}, and
\SIrange{6068}{6076}{\angstrom}. From the estimated S/N, ARES can give an
estimate on the very important \emph{rejt} parameters
\citep[see][for more information]{Sousa2015a}. After measuring the EWs with ARES,
we used the FASMA minimization routine described in Section~\ref{sub:EW_method}
to determine the stellar atmospheric parameters. The results are presented in
Figure~\ref{fig:FASMATest} which shows $T_\mathrm{eff}$, $\log g$,
[\ion{Fe}/\ion{H}], and $\xi_\mathrm{micro}$ for FASMA against those of
\citet{Sousa2011}.

The sample contains stars with $T_\mathrm{eff}$ too cold for the line list used.
As described in Section~\ref{sub:EW_method} we should then convert the line list
by \citet{Sousa2008a} to the line list presented in \citet{Tsantaki2013}.
However, since this line list was not available when \citet{Sousa2011} derived
parameters, we did not make this change in order to make a fair comparison for
FASMA.

The mean of the difference between parameters from \citet{Sousa2011} and those
by FASMA are presented in Table~\ref{tab:FASMATest}.

\begin{table}[htb!]
    \caption{Difference in derived parameters by \citet{Sousa2011}
    and FASMA. The second column is the mean difference with EWs measured by
    ARES in FASMA, while the third column is the mean difference using
    20 randomly stars with the exact same EWs from \citet{Sousa2011}.}
    \label{tab:FASMATest}
    \centering
    \begin{tabular}{lrr}
      \hline\hline
      Parameter             &  Mean difference         & Same line list        \\
      \hline
      $T_\mathrm{eff}$      &  $\SI{16(36)}{K}$        & $\SI{21(11)}{K}$      \\
      $\log g$              &  $\num{-0.04(7)}$        & $\num{-0.007(9)}$     \\
      $[\ion{Fe}/\ion{H}]$  &  $\num{0.03(2)}$         & $\num{0.004(9)}$      \\
      $\xi_\mathrm{micro}$  &  $\SI{-0.04(14)}{km/s}$  & $\SI{0.04(2)}{km/s}$  \\
      \hline
    \end{tabular}
\end{table}

The comparison is very consistent, as expected, and the small offsets are within
the errors except for metallicity. This can be due to different versions of
MOOG, measured line lists (i.e. using slightly different settings/version of
ARES to measure the EWs), interpolation of atmosphere grid, and minimization
routine. Most likely the difference will be due to the different \emph{rejt}
parameters used in ARES, which can alter the EWs systematically and hence the
metallicity. We therefore randomly selected 20 stars with different
$T_\mathrm{eff}$ and used the EWs directly from \citet{Sousa2011} to derive
parameters. The results are presented in the last column of
Table~\ref{tab:FASMATest}. We note that the $\log gf$ values from the original
line lists by \citet{Sousa2011}, which used the MOOG 2002 version, were not
changed for the 2014 version of MOOG. This might lead to some errors as well.
However, the offsets are very small and are compatible with the errors on
parameters normally obtained from high-quality spectra.

\subsection{Web interface}
\label{sub:Web interface}

We provide a web interface for FASMA. In the web interface it is possible to use
the line list provided with FASMA to measure the EWs of a spectrum. The spectrum
is expected to be in a 1D format with the wavelength information stored in the
header keys, CRVAL1 for the minimum wavelength, CDELT1 for the stepping in
wavelengths, and NAXIS1 for the number of points. This can be used for all the
available FASMA methods described above. ARES does the normalization, but the
best results are found when the spectrum is properly reduced (by removal of
cosmic rays, normalization, etc.).

The web interface can be found at the following link
\url{http://www.iastro.pt/fasma}, where more information is available on each of
the drivers. The user must provide an email. This is only used to send the
results after the calculations.

\section{SWEET-Cat update}

\subsection{Data}
\label{sec:data}
In this paper we derived parameters for a sample of 50 stars, 43 were observed
by our team using the UVES/VLT \citep{UVES}, FEROS/2.2m telescope in La Silla
\citep{FEROS}, and FIES/NOT \citep{FIES} spectrographs. The remaining spectra
(23) were found in various archives. We use spectra from the HARPS/3.6m
telescope in La Silla \citep{HARPS} and ESPaDOnS/CFHT \citep{ESPADONS}. Some
characteristics of the spectrographs are presented in
Table~\ref{tab:instruments} with the mean S/N for the spectra used. The S/N for
each star can be seen in Table~\ref{tab:results} along with the atmospheric
parameters of the stars. The S/N is measured automatically by ARES, but we note
that ARES smoothes the spectra before measuring the S/N, hence it is listed
higher than the actual S/N. These 50 stars are confirmed exoplanet hosts listed
in SWEET-Cat. However, they belonged to the list of stars that have not been
analysed by our team. We therefore increase the number of stars analysed in a
homogeneous way, which is the goal of SWEET-Cat.

We obtain the spectra with the highest possible resolution for a given
spectrograph, and in cases with multiple observations, we include all the
observations unless a spectrum is close to the saturation limit for a given
spectrograph. For multiple spectra, we combine them after first correcting the
radial velocity (RV) and using a sigma clipper to remove cosmic rays. The
individual spectra are then combined to a single spectrum for a given star to
increase the S/N. This single spectrum is used in the analysis described below.
For most of the spectra in the archive included here, several spectra were
combined as described above, while for the observations dedicated to this work,
the spectrum would be a single spectrum, or in cases of faint stars, it would be
observed a few times to reach the desired S/N. This is mostly due to the
difference in science cases behind the observations; for example,  the HARPS
spectra were used for RV monitoring or follow-up of the exoplanet(s), while the
UVES spectra were used to characterize stellar parameters.

\begin{table}[htb!]
    \caption{Spectrographs used for this paper with their spectral resolution,
             wavelength coverage, and mean S/N from the spectra used.}
    \label{tab:instruments}
    \centering
    \begin{tabular}{llll}
      \hline\hline
      Spectrograph & Resolution & Spectral range              &   Mean S/N  \\
      \hline
      HARPS        &    115 000 & $\SIrange{378}{691}{nm}$    &   642       \\
      UVES         &    110 000 & $\SIrange{480}{1100}{nm}$   &   212       \\
      ESPaDOnS     &     81 000 & $\SIrange{370}{1050}{nm}$   &   775       \\
      FIES         &     67 000 & $\SIrange{370}{730}{nm}$    &   763       \\
      FEROS        &     48 000 & $\SIrange{350}{920}{nm}$    &   208       \\
      \hline
    \end{tabular}
\end{table}

\subsection{Analysis}
\label{sec:results}
Here we present the sample of 50 stars. We were unable to derive parameters for
{16 of our targets (not included in the 50)}. For example, HD77065 is a
spectroscopic binary according to \cite{Pourbaix2004}, and the spectrum is
contaminated by the companion star. This make EW measurement very difficult,
hence we exclude it from the sample of collected spectra.

Moreover, we were not able to successfully derive parameters with this method
for Aldebaran, a well-known red giant star. Even though spectra of good quality
are available for a bright star like Aldebaran, this spectral type is
intrinsically difficult to analyse, due to the molecular absorption that arises
in the optical region at low $T_\mathrm{eff}$ (\SI{4055}{K} is listed in
SWEET-Cat as measured by \citealt{Hatzes2015}). The fact that we are not able to
derive parameters for Aldebaran is not a big concern since it has been well
studied with other techniques, and we can trust the parameters already listed in
SWEET-Cat.

In total, we removed 16 stars from our sample because the parameters for these
stars did not converge during the minimization procedure. The $T_\mathrm{eff}$
for these stars is either too hot, above \SI{7500}{K}, or too cold, below
\SI{4000}{K}, for the EW method to work.

The remaining 50 stars are presented in Table~\ref{tab:results}. We note that we
apply a correction to the spectroscopic $\log g$ based on asteroseismology as
found by \citet{Mortier2014}. We only use this correction for FGK dwarf stars,
i.e. between $\SI{4800}{K}\leq T_\mathrm{eff} \leq \SI{6500}{K}$ and $\log
g\geq4.2$. For stars with a $\log g$ lower than this limit we do not apply the
corrections, and if the $\log g$ changes to below this limit after the
correction, we go back to using the spectroscopic $\log g$ again. The correction
for $\log g$ depends on both $T_\mathrm{eff}$ and $\log g$. The correction can
be up to 0.5 dex, depending on the $T_\mathrm{eff}$.

\begin{figure}[tpb]
    \centering
    \includegraphics[width=1.0\linewidth,natwidth=440,natheight=290]{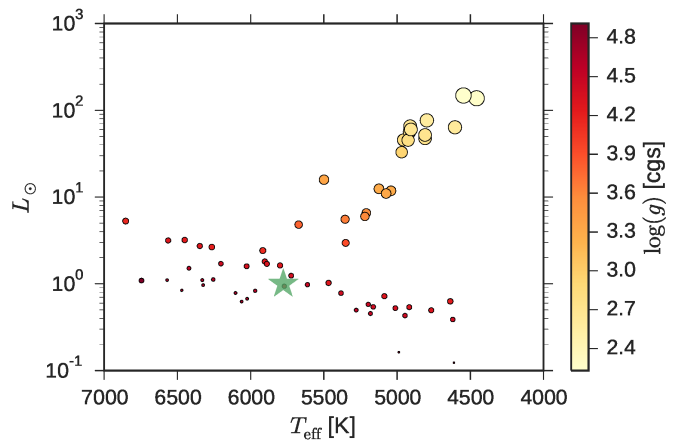}
    \caption{Hertzprung-Russel diagram of our sample with the Sun as a green
    star. The size of the points represents the $\log g$, with bigger points
    being smaller $\log g$ (giants), and vice versa. Red points are the dwarfs, while yellow points are the
    giants.}
    \label{fig:HRD}
\end{figure}

We present a Hertzprung-Russel diagram (HRD) of our sample in
Figure~\ref{fig:HRD}, which is made with a tool for post-processing the results
saved to a table by FASMA.

\begin{figure*}[tpb]
    \centering
    \includegraphics[width=1.0\linewidth,natwidth=870,natheight=580]{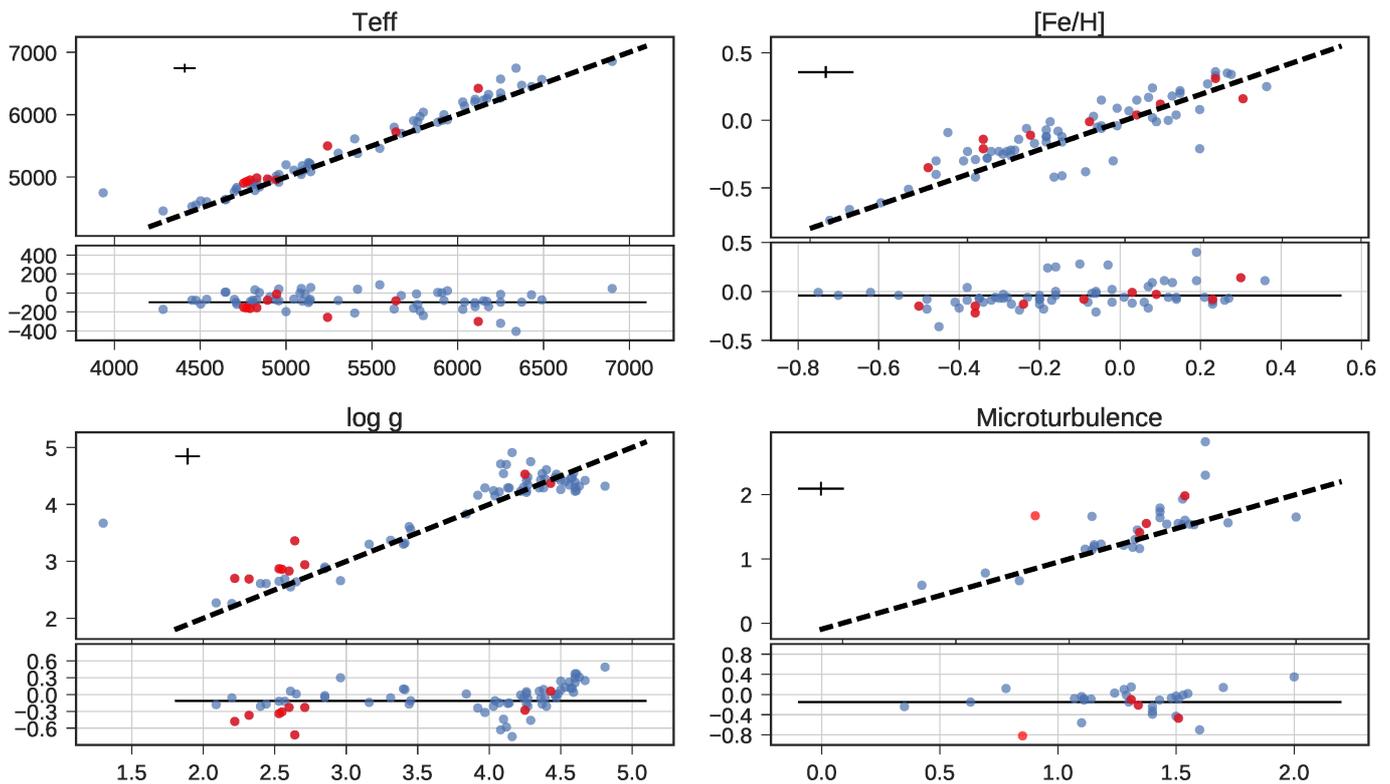}
    \caption{Atmospheric parameters of the updated planet host stars. The x-axis
    shows the previous values in SWEET-Cat (see Table~\ref{tab:oldSC}), while the
    y-axis indicates the new updated values. In the upper left corner of each of the
    four plots we show the typical error on the parameters. The red points
    are outliers, as discussed in Section~\ref{sec:Discussion}.}
    \label{fig:update}
\end{figure*}

The new atmospheric parameters are presented in Figure~\ref{fig:update} against
the literature values that we listed previously in SWEET-Cat. The red points
showing the location of the outliers as discussed below are mainly visible
outliers for $\log g$ at the low end, i.e. for the sub-giant to giant stars. The
old parameters are listed in Table~\ref{tab:oldSC}. Out of 2437 stars discovered
to be a planet hosts, 21\% have been analysed in the homogeneous way as
described in this work. We note that currently the limiting factor for
increasing the sample of stars analysed in the homogeneous way is the magnitude
of the planet hosts. Many planet hosts have been found through space missions
such as \emph{Kepler} and \emph{CoRoT} using the transiting method. Most of
these stars are faint and thus make them observationally expensive for the
spectroscopic analysis required here. For stars brighter than magnitude 12 the
completeness (i.e. the stars analysed in a homogeneous way compared to the ones
we have not analysed yet) is now up to 77\%, while it is at 85\% for exoplanet
hosts brighter than magnitude 10.

\begin{figure}[tpb]
    \centering
    \includegraphics[width=1.0\linewidth,natwidth=450,natheight=300]{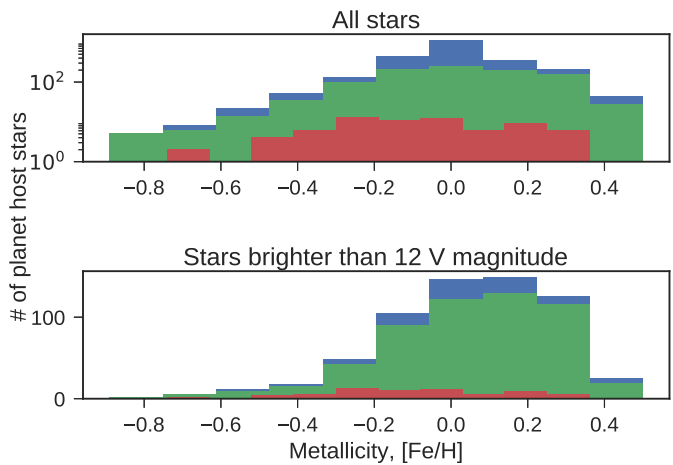}
    \caption{Metallicity distribution. Top: All
             stars (logarithmic scale) in SWEET-Cat, divided in three groups: blue
             (the largest distribution) are all stars available in SWEET-Cat;
             green (the middle distribution) are the stars with homogeneity flag
             1, i.e. analysed using the method described in this paper; and red
             (the smallest distribution) are all new stars added from this
             paper. Bottom: 12 V magnitude cut to exclude
             stars which are currently unavailable spectroscopically.}
    \label{fig:distribution}
\end{figure}

Understanding the metallicity distribution for all the planet host stars is
important in order to understand planet formation, for example. We present a
distribution in Figure~\ref{fig:distribution}. The sample is divided in two, for
all planet hosts in SWEET-Cat and for stars brighter than 12 V magnitude. Dimmer
stars are mainly observed with the \emph{Kepler} space mission. These dim stars
are very time consuming, and hence expensive to observe. Out of the 2437 stars
in SWEET-Cat, 664 are brighter than 12 V magnitude. We note that more than 1500
of the stars do not have a V magnitude. The majority are stars observed with
\emph{Kepler}. Our group have analysed 563 stars in SWEET-Cat including the 50
stars presented in this work.

\subsection{Discussion}
\label{sec:Discussion}
\begin{figure}[tpb]
    \centering
    \includegraphics[width=1.0\linewidth]{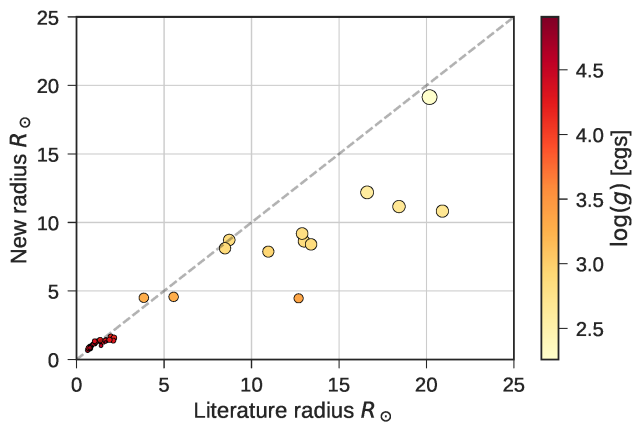}
    \caption{Stellar radius on both axes calculated based on \citet{Torres2010}.
    The x-axis shows the stellar radius based on the atmospheric parameters
    from the literature, while the y-axis indicates the new homogeneous parameters
    presented here. The colour and size indicate the surface gravity. This clearly
    shows that the disagreement is biggest for more evolved stars.}
    \label{fig:RR}
\end{figure}

We compute radius and mass of all the 50 new stars updated in SWEET-Cat (even
the ones whose parameters may not be reliable, in order to be complete) using
the empirical formula presented in \citet{Torres2010}. Some of the stars have
radii derived from different methods, usually from isochrones. These radii
generally show a good correlation with radii derived from \citet{Torres2010} if
the literature parameters of $T_\mathrm{eff}$, $\log g$, and
$[\ion{Fe}/\ion{H}]$ are used. However, if we make a comparison with the new
radius derived using the parameters presented here, the results can differ by up
to 65\%. We show in Figure~\ref{fig:RR} how the radius calculated from
\citet{Torres2010} differs between the literature atmospheric parameters and the
new homogeneous atmospheric parameters presented here. We note that stellar
radii are provided by many of the authors from different discovery papers, but
we chose to compare the atmospheric parameters via the derivation of the stellar
radius, as described above, rather than comparing the stellar radii from
different methods.

In the sections below we discuss the systems (seven stars, eight exoplanets)
where the radius or mass of the stars changes more than 25\% and how this
influences the planetary parameters. The changes in radius for a star is
primarily due to changes in $\log g$, which can be used as an indicator of the
evolutionary stage of a star.

We rederive the planetary radius, mass, and semi-major axis when possible
following the three simple scaling relations based on Newton's law of gravity
\citep{Newton1687} for deriving mass and distance and simple geometry for radius
\citep[see e.g.][]{Torres2008}

\begin{align}
    M_\mathrm{pl,new} &= \left(\frac{M_\mathrm{\ast,lit}}{M_\mathrm{\ast,new}}\right)^{-2/3} M_\mathrm{pl,lit}  \\
    R_\mathrm{pl,new} &= \left(\frac{R_\mathrm{\ast,lit}}{R_\mathrm{\ast,new}}\right) R_\mathrm{pl,lit} \\
    a_\mathrm{pl,new} &= \left(\frac{M_\mathrm{\ast,lit}}{M_\mathrm{\ast,new}}\right)^{1/3} a_\mathrm{pl,lit},
\end{align}
where the subscript ``lit'' denotes the value from the literature used in the
comparison, the subscript ``new'' indicates the new computed values, the
subscript ``pl'' is short for planet, and the subscript ``$\ast$'' is short for
star; $M$, $R$, and $a$ are mass, radius, and semi-major axis, respectively. We
note that for the literature values, we use the values reported directly from
the literature and not the derived radius and mass from \citet{Torres2010}. To
identify outliers, we compare radii and masses derived from \citet{Torres2010}
since this is a measure of how the atmospheric parameters have changed.

\subsubsection{HAT-P-46}
\label{sub:HAT-P-46}
HAT-P-46 has two known exoplanets according to \citet{Hartmann2014}. The outer
planet HAT-P-46 c is not transiting, hence we do not have a radius for this
planet. The results we present in this paper for this star come from UVES/VLT
data with a S/N of 208. \citet{Hartmann2014} derives the following spectroscopic
parameters: $T_\mathrm{eff}=\SI{6120(100)}{K}$, $\log g=4.25\pm0.11$, and
$[\ion{Fe}/\ion{H}]=0.30\pm0.10$. We note that for this star the asteroseismic
correction we apply (see Section~\ref{sec:results}) results in a corrected $\log
g$ below 4.2dex, so we used the spectroscopic $\log g$ for this star.

If we derive the mass and radius of HAT-P-46 b with our new parameters, we
obtain $R_\mathrm{pl} = 0.93R_J$, while \citet{Hartmann2014} derived
$R_\mathrm{pl} = 1.28R_J$. We see no change in mass (\citealt{Hartmann2014}
found $M_\mathrm{pl}=0.49M_J$); however, there is a decrease in the radius, and
we end up with a more dense planet, $\rho_\mathrm{pl}=\SI{0.76}{g/cm^3}$ from
$\rho_\mathrm{pl}=\SI{0.28\pm0.10}{g/cm^3}$.

As the secondary companion does not transit we only have a limit on the minimum
mass for this planet. Here we get $M\sin i_\mathrm{pl} = 1.97M_J$ and
\citet{Hartmann2014} presented $M\sin i_\mathrm{pl} = 2.00M_J$, so a very small
change, as expected.

\subsubsection{HD 120084}
\label{sub:HD_120084}
The exoplanet orbiting this star with a period of 2082 days and a quite
eccentric orbit at 0.66 was discovered by \citet{Sato2013}. The atmospheric
parameters were derived by \citet{Takeda2008} using a similar method to that
described in this paper. The quality of the spectra they analysed, however, were
not as high as those used here. Using the HIDES spectrograph at the 188cm
reflector at NAOJ, \citet{Takeda2008} reported an average S/N for their sample
of 100-300 objects at a resolving power of 67 000. We used data from ESPaDOnS
with a resolving power of 81 000, and with a S/N for this star of 850. With our
new parameters we obtain a slightly lower stellar mass for the star at
$1.93M_\odot$ compared to $2.39M_\odot$ obtained by \cite{Takeda2008}, hence the
minimum planetary mass is also slightly lower, from $m_\mathrm{pl}\sin i=4.5M_J$
to $m_\mathrm{pl}\sin i=3.9M_J$. We see a 28\% decrease in the stellar radius,
from $9.12R_\odot$ to $7.81R_\odot$. Since there are no observations of the
planet transiting, the planetary radius has not been computed.

\subsubsection{HD 233604}
\label{sub:HD_233604}
HD 233604 b was discovered by \citet{Nowak2013}, while the atmospheric
parameters of the star were derived by \citet{Zielinski2012}, who used the same
method as described in this paper using the HRS spectrograph at HET with a
resolving power of 60 000 with a typical S/N at 200-250. We obtained the
spectrum for this star using the FIES spectrograph with a slightly higher
resolution at 67 000, and similar but also slightly higher S/N at 320 for this
star.

This planet is in a very close orbit with a semi-major axis of $\sim 15R_\ast$
($R_\ast$ is the stellar radius) using the parameters from \citet{Nowak2013}.
Using the updated parameters presented in this paper we see a slight increase in
the stellar mass from $1.5M_\odot$ to $1.9M_\odot$, and a decrease in stellar
radius from $10.5R_\odot$ to $8.6R_\odot$. This increases the semi-major axis to
$\sim 21R_\ast$. We note that the correct stellar radii are used to describe the
semi-major axis in both cases. The increase in stellar mass leads to an increase
in the minimum planetary mass, from $m_\mathrm{pl}\sin i=6.58M_J$ to
$m_\mathrm{pl}\sin i=7.79M_J$.

\citet{Nowak2013} found a high \ion{Li}{} abundance at
$A(\ion{Li}{})_\mathrm{LTE}=1.400\pm0.042$ for this star and speculated that
this star might have engulfed a planet. A more likely explanation is that this
star has not yet reached the first dredge-up process \citep{Nowak2013}. We found
a much lower value, $A(\ion{Li}{})_\mathrm{LTE}=0.92$, and hence do not find the
star to be \ion{Li}{} rich. The \ion{Li}{} abundance we find is in excellent
agreement with \citet{Adamow2014}. Even applying a NLTE correction, as was done
in \citet{Adamow2014} ($A(\ion{Li}{})_\mathrm{NLTE}=1.08$), this star is not
\ion{Li}{} rich.

\subsubsection{HD 5583}
\label{sub:HD_5583}
This exoplanet was discovered by \citet{Niedzielski2016} with an orbital period
of 139 days around a K giant. This exoplanet was discovered with the radial
velocity technique, and we do not have a planetary radius. The stellar
parameters were derived in a similar manner to that presented here
\citep[see][and references therein]{Niedzielski2016}; our biggest disagreement
is in the surface gravity. We derive a $\log g$ that is higher by 0.34 dex,
which gives a stellar radius that is smaller by 37\%. The derived mass is 15\%
higher, which in turn increases the minimum planetary mass from
$m_\mathrm{pl}\sin i=5.78M_J$ to $m_\mathrm{pl}\sin i=8.63M_J$. Even with the
increase in mass, it is still within the planetary regime for most inclinations,
as was noted by \citet{Niedzielski2016}.

\subsubsection{HD 81688}
\label{sub:HD81688}
This exoplanet was discovered by \citet{Sato2008} with the RV method. The host
star is a metal-poor K giant. The atmospheric parameters presented in
\citet{Sato2008} are obtained via the same method as presented in this paper,
and the agreement is quite good. Once again the big disagreement is in the
surface gravity: ours is 0.48 dex higher. Even though the stellar parameters,
and hence the planetary parameters, do change, the radius and mass we derive are
not far from the values presented in the paper by \citet{Sato2008}. This is a
case where the star was marked as an outlier, due to the comparison between the
radius and mass derived from \citet{Torres2010}.

The new stellar mass is the same as before, $2.1M_\odot$. The stellar radius
changed from $13.0R_\odot$ to $10.8R_\odot$. Since a transit of this star has
not been observed and the stellar mass remains the same, we do not see any
change in the planetary parameters.

We note that this system is in an interesting configuration with a very close
orbit around an evolved star. This system, among others, has been the subject of
work on planet engulfment \citep[see e.g.][]{Kunitomo2011}.

\subsubsection{HIP 107773}
\label{sub:HIP_107773}
The planetary companion was presented in \citet{Jones2015} as an exoplanet
around an intermediate-mass evolved star. The stellar parameters were obtained
from the analysis by \citet{Jones2011} using the same method as presented here,
but with a different line list, which might lead to some disagreements. For this
star we derive a higher $\log g$ (2.83 dex compared to 2.60 dex), thus we derive
a slightly smaller star with $11.6R_\odot$ to $9.2R_\odot$ and $2.4M_\odot$ to
$2.1M_\odot$ for radius and mass of the star, respectively. The other
atmospheric parameters are very similar to those derived by \citet{Jones2011}.
This leads to a reduced minimum mass of the planetary companion from $m\sin
i=1.98M_J$ to $m\sin i=1.78M_J$. The planetary radius has not been measured.

\subsubsection{WASP-97}
\label{sub:WASP-97}
The exoplanet orbiting WASP-97 was discovered by \citet{Hellier2014}. The host
star parameters were derived using a similar method to that described in this
paper after co-adding several spectra from the CORALIE spectrograph. They reach
a S/N of 100 with a spectral resolution of 50 000. The parameters presented here
come from the UVES spectrograph with a S/N of more than 200.

The parameters do not change much for this planet. The planetary mass changes
from $m_\mathrm{pl}\sin i=1.32M_J$ to $m_\mathrm{pl}\sin i=1.37M_J$ and the
radius from $1.13R_J$ to $1.42R_J$. This affects the density quite strongly; it
changes from $\SI{1.13}{g/cm^3}$ to $\SI{0.59}{g/cm^3}$. This exoplanet is then
in the same category as Saturn; its density is lower than water, but it is
slightly larger than Jupiter.

\subsubsection{$\omega$ Serpentis (ome Ser)}
\label{sub:ome_Ser}
The exoplanet orbiting this star with a period of 277 days and an eccentric
orbit at 0.11 was also presented by \citet{Sato2013}. The atmospheric parameters
were derived in the same way as for HD 120084. We used data from FIES with a
resolving power of 67 000, and with a S/N for this star of 1168. With our new
parameters we obtain a slightly higher stellar mass for the star at
$2.19M_\odot$ compared to the value of $2.17M_\odot$ obtained by
\cite{Takeda2008}. This change is not significant enough to change the minimum
planetary mass at $m_\mathrm{pl}\sin i=1.7M_J$. The stellar radius decreases by
more than one solar radius, from $12.3R_\odot$ to $11.1R_\odot$. However, since
there are no observations of transiting exoplanets, we cannot see the change in
the planetary radius.

\subsubsection{o Ursa Major (omi UMa)}
\label{sub:omiUMa}
omi UMa b was discovered by \citet{Sato2012} using the RV method. The stellar
parameters are from \citet{Takeda2008}, as discussed above. The spectrum used
for this star is from ESPaDOnS with a S/N of more than 500 compared to the value
of 100-300 reached for the large sample presented in \citet{Takeda2008}. The
luminosity and mass for omi UMa were obtained from theoretical evolutionary
tracks \citep[see][and references therein]{Sato2012}. The radius was then
estimated using the Stefan-Boltzmann relationship, using the measured luminosity
and $T_\mathrm{eff}$.

The parameters presented here mainly differ in the surface gravity: ours is 0.72
dex higher at $\log g=3.36$. This leads to a big change in stellar mass and
radius from $3.1M_\odot$ to $1.6M_\odot$ and $14.1R_\odot$ to $4.5R_\odot$,
respectively. \citet{Sato2012} have reported that omi UMa b is the first planet
candidate around a star more massive than $3M_\odot$. With these updated
results, the minimum mass of the planet is now $m\sin i=2.7M_J$, whereas
previously it was $m\sin i=4.1M_J$ \citep{Sato2012}. The exoplanet is not
reported to transit, as seen from Earth, so we do not have a radius for this
exoplanet, which would have changed a great deal with these new results.

\section{Conclusion}
\label{sec:conclusion}

With this update we bring the completeness of SWEET-Cat for stars brighter than
magnitude 10 (V band) up to 85\% (77\% for stars brighter than 12). The
parameters are continuously updated and are available for the public in an
easily accessibly
form\footnote{\url{https://www.astro.up.pt/resources/sweet-cat/}}. The
importance of the homogeneous analysis which we keep striving for is shown in
Figure~\ref{fig:RR} where we see quite different derived stellar radii when the
atmospheric parameters are obtained via different methods. Even using the same
method but with a different setup (different line list, minimization routine,
atmospheric models, etc.), we can arrive at different results. This again shows
the importance of analysing the stars in a homogeneous way. As has been
discussed in Section~\ref{sec:Discussion} this has a direct impact on the
planetary parameters. It is of great importance to know the planetary parameters
very well, for individual systems and also for an ensemble. With accurate and
precise planetary parameters we will be able to distinguish the different
possible compositions, whether gas giants, water worlds, or rocky planets.

Finally we also provide an online tool for deriving the stellar atmospheric
parameters using FASMA. We recommend this tool only for spectra and stars where
this method is working, i.e. high-resolution spectra with a high S/N. The stars
can be FGK dwarfs and FGK subgiants/giants. We are working on applying this
method to the near-IR in order to include the cool solar-type stars.

\begin{acknowledgements}

This work was supported by Funda\c{c}\~ao para a Ci\^encia e a Tecnologia (FCT)
through national funds and by FEDER through COMPETE2020 by these grants
UID/FIS/0Q4434/2013 \& POCI-01-0145-FEDER-007672, PTDC/FIS-AST/7073/2014 \&
POCI-01-0145-FEDER-016880 and PTDC/FIS-AST/1526/2014 \& POCI-01-0145-FEDER-016886.

S.G.S. and N.C.S. acknowledge the support from FCT through the Investigador FCT
Contracts of reference IF/00028/2014/CP1215/CT0002 and
IF/00169/2012/CP0150/CT0002, respectively, and POPH/FSE (EC) by FEDER funding
through the program “Programa Operacional de Factores de Competitividade -
COMPETE”.

G.D.C.T. was supported by the PhD fellowship PD/BD/113478/2015funded by FCT
(Portugal) and POPH/FSE (EC).

E.D.M. acknowledge the support by the fellowship SFRH/BPD/76606/2011funded by
FCT (Portugal) and POPH/FSE (EC).

A.C.S.F. was supported by grant 234989/2014-9 from CNPq (Brazil).

A.M. received funding from the European Union Seventh Framework Programme
(FP7/2007-2013) under grant agreement number 313014 (ETAEARTH).

This research has made use of the SIMBAD database operated at CDS, Strasbourg
(France).

We thank the anonymous referee for the useful comments and suggestions which
helped clarify the manuscript.

\end{acknowledgements}

\bibliographystyle{aa}
\bibliography{thesis}

\begin{thebibliography}{81}
\expandafter\ifx\csname natexlab\endcsname\relax\def\natexlab#1{#1}\fi

\bibitem[{{Adam{\'o}w} {et~al.}(2014){Adam{\'o}w}, {Niedzielski}, {Villaver},
  {Wolszczan}, \& {Nowak}}]{Adamow2014}
{Adam{\'o}w}, M., {Niedzielski}, A., {Villaver}, E., {Wolszczan}, A., \&
  {Nowak}, G. 2014, \aap, 569, A55

\bibitem[{{Adibekyan} {et~al.}(2015{\natexlab{a}}){Adibekyan}, {Figueira},
  {Santos}, {Sousa}, {Faria}, {Delgado-Mena}, {Oshagh}, {Tsantaki}, {Hakobyan},
  {Gonz{\'a}lez Hern{\'a}ndez}, {Su{\'a}rez-Andr{\'e}s}, \&
  {Israelian}}]{Adibekyan2015b}
{Adibekyan}, V., {Figueira}, P., {Santos}, N.~C., {et~al.} 2015{\natexlab{a}},
  \aap, 583, A94

\bibitem[{{Adibekyan} {et~al.}(2015{\natexlab{b}}){Adibekyan}, {Benamati},
  {Santos}, {Alves}, {Lovis}, {Udry}, {Israelian}, {Sousa}, {Tsantaki},
  {Mortier}, {Sozzetti}, \& {De Medeiros}}]{Adibekyan2015}
{Adibekyan}, V.~Z., {Benamati}, L., {Santos}, N.~C., {et~al.}
  2015{\natexlab{b}}, \mnras, 450, 1900

\bibitem[{{Adibekyan} {et~al.}(2012){Adibekyan}, {Sousa}, {Santos}, {Delgado
  Mena}, {Gonz{\'a}lez Hern{\'a}ndez}, {Israelian}, {Mayor}, \&
  {Khachatryan}}]{Adibekyan2012}
{Adibekyan}, V.~Z., {Sousa}, S.~G., {Santos}, N.~C., {et~al.} 2012, \aap, 545,
  A32

\bibitem[{{Anders} \& {Grevesse}(1989)}]{Anders1989}
{Anders}, E. \& {Grevesse}, N. 1989, \gca, 53, 197

\bibitem[{{Anderson} {et~al.}(2012){Anderson}, {Collier Cameron}, {Gillon},
  {Hellier}, {Jehin}, {Lendl}, {Maxted}, {Queloz}, {Smalley}, {Smith},
  {Triaud}, {West}, {Pepe}, {Pollacco}, {S{\'e}gransan}, {Todd}, \&
  {Udry}}]{Anderson2012}
{Anderson}, D.~R., {Collier Cameron}, A., {Gillon}, M., {et~al.} 2012, \mnras,
  422, 1988

\bibitem[{{Andreasen} {et~al.}(2016){Andreasen}, {Sousa}, {Delgado Mena},
  {Santos}, {Tsantaki}, {Rojas-Ayala}, \& {Neves}}]{Andreasen2016}
{Andreasen}, D.~T., {Sousa}, S.~G., {Delgado Mena}, E., {et~al.} 2016, \aap,
  585, A143

\bibitem[{{Barclay} {et~al.}(2013){Barclay}, {Rowe}, {Lissauer}, {Huber},
  {Fressin}, {Howell}, {Bryson}, {Chaplin}, {D{\'e}sert}, {Lopez}, {Marcy},
  {Mullally}, {Ragozzine}, {Torres}, {Adams}, {Agol}, {Barrado}, {Basu},
  {Bedding}, {Buchhave}, {Charbonneau}, {Christiansen},
  {Christensen-Dalsgaard}, {Ciardi}, {Cochran}, {Dupree}, {Elsworth},
  {Everett}, {Fischer}, {Ford}, {Fortney}, {Geary}, {Haas}, {Handberg},
  {Hekker}, {Henze}, {Horch}, {Howard}, {Hunter}, {Isaacson}, {Jenkins},
  {Karoff}, {Kawaler}, {Kjeldsen}, {Klaus}, {Latham}, {Li}, {Lillo-Box},
  {Lund}, {Lundkvist}, {Metcalfe}, {Miglio}, {Morris}, {Quintana}, {Stello},
  {Smith}, {Still}, \& {Thompson}}]{Barclay2013}
{Barclay}, T., {Rowe}, J.~F., {Lissauer}, J.~J., {et~al.} 2013, \nat, 494, 452

\bibitem[{{Bedell} {et~al.}(2015){Bedell}, {Mel{\'e}ndez}, {Bean},
  {Ram{\'{\i}}rez}, {Asplund}, {Alves-Brito}, {Casagrande}, {Dreizler},
  {Monroe}, {Spina}, \& {Tucci Maia}}]{Bedell2015}
{Bedell}, M., {Mel{\'e}ndez}, J., {Bean}, J.~L., {et~al.} 2015, \aap, 581, A34

\bibitem[{{Boisse} {et~al.}(2013){Boisse}, {Hartman}, {Bakos}, {Penev},
  {Csubry}, {B{\'e}ky}, {Latham}, {Bieryla}, {Torres}, {Kov{\'a}cs},
  {Buchhave}, {Hansen}, {Everett}, {Esquerdo}, {Szklen{\'a}r}, {Falco},
  {Shporer}, {Fulton}, {Noyes}, {Stefanik}, {L{\'a}z{\'a}r}, {Papp}, \&
  {S{\'a}ri}}]{Boisse2013}
{Boisse}, I., {Hartman}, J.~D., {Bakos}, G.~{\'A}., {et~al.} 2013, \aap, 558,
  A86

\bibitem[{{Brucalassi} {et~al.}(2014){Brucalassi}, {Pasquini}, {Saglia},
  {Ruiz}, {Bonifacio}, {Bedin}, {Biazzo}, {Melo}, {Lovis}, \&
  {Randich}}]{Brucalassi2014}
{Brucalassi}, A., {Pasquini}, L., {Saglia}, R., {et~al.} 2014, \aap, 561, L9

\bibitem[{{Bryan} {et~al.}(2012){Bryan}, {Alsubai}, {Latham}, {Parley},
  {Collier Cameron}, {Quinn}, {Carter}, {Fulton}, {Berlind}, {Brown},
  {Buchhave}, {Calkins}, {Esquerdo}, {F{\H u}r{\'e}sz}, {Gr{\aa}e
  J{\o}rgensen}, {Horne}, {Stefanik}, {Street}, {Torres}, {West}, {Dominik},
  {Harps{\o}e}, {Liebig}, {Calchi Novati}, {Ricci}, \& {Skottfelt}}]{Bryan2012}
{Bryan}, M.~L., {Alsubai}, K.~A., {Latham}, D.~W., {et~al.} 2012, \apj, 750, 84

\bibitem[{{Campante} {et~al.}(2015){Campante}, {Barclay}, {Swift}, {Huber},
  {Adibekyan}, {Cochran}, {Burke}, {Isaacson}, {Quintana}, {Davies}, {Silva
  Aguirre}, {Ragozzine}, {Riddle}, {Baranec}, {Basu}, {Chaplin},
  {Christensen-Dalsgaard}, {Metcalfe}, {Bedding}, {Handberg}, {Stello},
  {Brewer}, {Hekker}, {Karoff}, {Kolbl}, {Law}, {Lundkvist}, {Miglio}, {Rowe},
  {Santos}, {Van Laerhoven}, {Arentoft}, {Elsworth}, {Fischer}, {Kawaler},
  {Kjeldsen}, {Lund}, {Marcy}, {Sousa}, {Sozzetti}, \& {White}}]{Campante2015}
{Campante}, T.~L., {Barclay}, T., {Swift}, J.~J., {et~al.} 2015, \apj, 799, 170

\bibitem[{{Collins} {et~al.}(2014){Collins}, {Eastman}, {Beatty}, {Siverd},
  {Gaudi}, {Pepper}, {Kielkopf}, {Johnson}, {Howard}, {Fischer}, {Manner},
  {Bieryla}, {Latham}, {Fulton}, {Gregorio}, {Buchhave}, {Jensen}, {Stassun},
  {Penev}, {Crepp}, {Hinkley}, {Street}, {Cargile}, {Mack}, {Oberst}, {Avril},
  {Mellon}, {McLeod}, {Penny}, {Stefanik}, {Berlind}, {Calkins}, {Mao},
  {Richert}, {DePoy}, {Esquerdo}, {Gould}, {Marshall}, {Oelkers}, {Pogge},
  {Trueblood}, \& {Trueblood}}]{Collins2014}
{Collins}, K.~A., {Eastman}, J.~D., {Beatty}, T.~G., {et~al.} 2014, \aj, 147,
  39

\bibitem[{{Dekker} {et~al.}(2000){Dekker}, {D'Odorico}, {Kaufer}, {Delabre}, \&
  {Kotzlowski}}]{UVES}
{Dekker}, H., {D'Odorico}, S., {Kaufer}, A., {Delabre}, B., \& {Kotzlowski}, H.
  2000, in \procspie, Vol. 4008, Optical and IR Telescope Instrumentation and
  Detectors, ed. M.~{Iye} \& A.~F. {Moorwood}, 534--545

\bibitem[{{Delrez} {et~al.}(2014){Delrez}, {Van Grootel}, {Anderson},
  {Collier-Cameron}, {Doyle}, {Fumel}, {Gillon}, {Hellier}, {Jehin}, {Lendl},
  {Neveu-VanMalle}, {Maxted}, {Pepe}, {Pollacco}, {Queloz}, {S{\'e}gransan},
  {Smalley}, {Smith}, {Southworth}, {Triaud}, {Udry}, \& {West}}]{Delrez2014}
{Delrez}, L., {Van Grootel}, V., {Anderson}, D.~R., {et~al.} 2014, \aap, 563,
  A143

\bibitem[{{Donati}(2003)}]{ESPADONS}
{Donati}, J.-F. 2003, in Astronomical Society of the Pacific Conference Series,
  Vol. 307, Solar Polarization, ed. J.~{Trujillo-Bueno} \& J.~{Sanchez
  Almeida}, 41

\bibitem[{{Frandsen} \& {Lindberg}(1999)}]{FIES}
{Frandsen}, S. \& {Lindberg}, B. 1999, in Astrophysics with the NOT, ed.
  H.~{Karttunen} \& V.~{Piirola}, 71

\bibitem[{{Gillon} {et~al.}(2013){Gillon}, {Anderson}, {Collier-Cameron},
  {Doyle}, {Fumel}, {Hellier}, {Jehin}, {Lendl}, {Maxted}, {Montalb{\'a}n},
  {Pepe}, {Pollacco}, {Queloz}, {S{\'e}gransan}, {Smith}, {Smalley},
  {Southworth}, {Triaud}, {Udry}, \& {West}}]{Gillon2013}
{Gillon}, M., {Anderson}, D.~R., {Collier-Cameron}, A., {et~al.} 2013, \aap,
  552, A82

\bibitem[{{G{\'o}mez Maqueo Chew} {et~al.}(2013){G{\'o}mez Maqueo Chew},
  {Faedi}, {Pollacco}, {Brown}, {Doyle}, {Collier Cameron}, {Gillon}, {Lendl},
  {Smalley}, {Triaud}, {West}, {Wheatley}, {Busuttil}, {Liebig}, {Anderson},
  {Armstrong}, {Barros}, {Bento}, {Bochinski}, {Burwitz}, {Delrez}, {Enoch},
  {Fumel}, {Haswell}, {H{\'e}brard}, {Hellier}, {Holmes}, {Jehin}, {Kolb},
  {Maxted}, {McCormac}, {Miller}, {Norton}, {Pepe}, {Queloz},
  {Rodr{\'{\i}}guez}, {S{\'e}gransan}, {Skillen}, {Stassun}, {Udry}, \&
  {Watson}}]{Gomez2013}
{G{\'o}mez Maqueo Chew}, Y., {Faedi}, F., {Pollacco}, D., {et~al.} 2013, \aap,
  559, A36

\bibitem[{{Gonzalez} \& {Laws}(2000)}]{Gonzalez2000}
{Gonzalez}, G. \& {Laws}, C. 2000, \aj, 119, 390

\bibitem[{{Gray}(2005)}]{Gray2006}
{Gray}, D.~F. 2005, {The Observation and Analysis of Stellar Photospheres, 3rd
  ed.}

\bibitem[{{Gustafsson} {et~al.}(2008){Gustafsson}, {Edvardsson}, {Eriksson},
  {J{\o}rgensen}, {Nordlund}, \& {Plez}}]{Gustafson2008}
{Gustafsson}, B., {Edvardsson}, B., {Eriksson}, K., {et~al.} 2008, \aap, 486,
  951

\bibitem[{{Hartman} {et~al.}(2012){Hartman}, {Bakos}, {B{\'e}ky}, {Torres},
  {Latham}, {Csubry}, {Penev}, {Shporer}, {Fulton}, {Buchhave}, {Johnson},
  {Howard}, {Marcy}, {Fischer}, {Kov{\'a}cs}, {Noyes}, {Esquerdo}, {Everett},
  {Szklen{\'a}r}, {Quinn}, {Bieryla}, {Knox}, {Hinz}, {Sasselov}, {F{\H
  u}r{\'e}sz}, {Stefanik}, {L{\'a}z{\'a}r}, {Papp}, \&
  {S{\'a}ri}}]{Hartman2012}
{Hartman}, J.~D., {Bakos}, G.~{\'A}., {B{\'e}ky}, B., {et~al.} 2012, \aj, 144,
  139

\bibitem[{{Hartman} {et~al.}(2014{\natexlab{a}}){Hartman}, {Bakos}, {Torres},
  {Kov{\'a}cs}, {Johnson}, {Howard}, {Marcy}, {Latham}, {Bieryla}, {Buchhave},
  {Bhatti}, {B{\'e}ky}, {Csubry}, {Penev}, {de Val-Borro}, {Noyes}, {Fischer},
  {Esquerdo}, {Everett}, {Szklen{\'a}r}, {Zhou}, {Bayliss}, {Shporer},
  {Fulton}, {Sanchis-Ojeda}, {Falco}, {L{\'a}z{\'a}r}, {Papp}, \&
  {S{\'a}ri}}]{Hartmann2014}
{Hartman}, J.~D., {Bakos}, G.~{\'A}., {Torres}, G., {et~al.}
  2014{\natexlab{a}}, \aj, 147, 128

\bibitem[{{Hartman} {et~al.}(2014{\natexlab{b}}){Hartman}, {Bakos}, {Torres},
  {Kov{\'a}cs}, {Johnson}, {Howard}, {Marcy}, {Latham}, {Bieryla}, {Buchhave},
  {Bhatti}, {B{\'e}ky}, {Csubry}, {Penev}, {de Val-Borro}, {Noyes}, {Fischer},
  {Esquerdo}, {Everett}, {Szklen{\'a}r}, {Zhou}, {Bayliss}, {Shporer},
  {Fulton}, {Sanchis-Ojeda}, {Falco}, {L{\'a}z{\'a}r}, {Papp}, \&
  {S{\'a}ri}}]{Hartman2014}
{Hartman}, J.~D., {Bakos}, G.~{\'A}., {Torres}, G., {et~al.}
  2014{\natexlab{b}}, \aj, 147, 128

\bibitem[{{Hatzes} {et~al.}(2015){Hatzes}, {Cochran}, {Endl}, {Guenther},
  {MacQueen}, {Hartmann}, {Zechmeister}, {Han}, {Lee}, {Walker}, {Yang},
  {Larson}, {Kim}, {Mkrtichian}, {D{\"o}llinger}, {Simon}, \&
  {Girardi}}]{Hatzes2015}
{Hatzes}, A.~P., {Cochran}, W.~D., {Endl}, M., {et~al.} 2015, \aap, 580, A31

\bibitem[{{H{\'e}brard} {et~al.}(2016){H{\'e}brard}, {Arnold}, {Forveille},
  {Correia}, {Laskar}, {Bonfils}, {Boisse}, {D{\'{\i}}az}, {Hagelberg},
  {Sahlmann}, {Santos}, {Astudillo-Defru}, {Borgniet}, {Bouchy}, {Bourrier},
  {Courcol}, {Delfosse}, {Deleuil}, {Demangeon}, {Ehrenreich}, {Gregorio},
  {Jovanovic}, {Labrevoir}, {Lagrange}, {Lovis}, {Lozi}, {Moutou},
  {Montagnier}, {Pepe}, {Rey}, {Santerne}, {S{\'e}gransan}, {Udry},
  {Vanhuysse}, {Vigan}, \& {Wilson}}]{Hebrard2016}
{H{\'e}brard}, G., {Arnold}, L., {Forveille}, T., {et~al.} 2016, \aap, 588,
  A145

\bibitem[{{H{\'e}brard} {et~al.}(2013){H{\'e}brard}, {Collier Cameron},
  {Brown}, {D{\'{\i}}az}, {Faedi}, {Smalley}, {Anderson}, {Armstrong},
  {Barros}, {Bento}, {Bouchy}, {Doyle}, {Enoch}, {G{\'o}mez Maqueo Chew},
  {H{\'e}brard}, {Hellier}, {Lendl}, {Lister}, {Maxted}, {McCormac}, {Moutou},
  {Pollacco}, {Queloz}, {Santerne}, {Skillen}, {Southworth}, {Tregloan-Reed},
  {Triaud}, {Udry}, {Vanhuysse}, {Watson}, {West}, \& {Wheatley}}]{Hebrard2013}
{H{\'e}brard}, G., {Collier Cameron}, A., {Brown}, D.~J.~A., {et~al.} 2013,
  \aap, 549, A134

\bibitem[{{Hellier} {et~al.}(2014){Hellier}, {Anderson}, {Cameron}, {Delrez},
  {Gillon}, {Jehin}, {Lendl}, {Maxted}, {Pepe}, {Pollacco}, {Queloz},
  {S{\'e}gransan}, {Smalley}, {Smith}, {Southworth}, {Triaud}, {Udry}, \&
  {West}}]{Hellier2014}
{Hellier}, C., {Anderson}, D.~R., {Cameron}, A.~C., {et~al.} 2014, \mnras, 440,
  1982

\bibitem[{{Hellier} {et~al.}(2012){Hellier}, {Anderson}, {Collier Cameron},
  {Doyle}, {Fumel}, {Gillon}, {Jehin}, {Lendl}, {Maxted}, {Pepe}, {Pollacco},
  {Queloz}, {S{\'e}gransan}, {Smalley}, {Smith}, {Southworth}, {Triaud},
  {Udry}, \& {West}}]{Hellier2012}
{Hellier}, C., {Anderson}, D.~R., {Collier Cameron}, A., {et~al.} 2012, \mnras,
  426, 739

\bibitem[{{Howard} {et~al.}(2011){Howard}, {Johnson}, {Marcy}, {Fischer},
  {Wright}, {Henry}, {Isaacson}, {Valenti}, {Anderson}, \&
  {Piskunov}}]{Howard2011}
{Howard}, A.~W., {Johnson}, J.~A., {Marcy}, G.~W., {et~al.} 2011, \apj, 730, 10

\bibitem[{{Johnson} {et~al.}(2011){Johnson}, {Clanton}, {Howard}, {Bowler},
  {Henry}, {Marcy}, {Crepp}, {Endl}, {Cochran}, {MacQueen}, {Wright}, \&
  {Isaacson}}]{Johnson2011}
{Johnson}, J.~A., {Clanton}, C., {Howard}, A.~W., {et~al.} 2011, \apjs, 197, 26

\bibitem[{{Jones} {et~al.}(2011){Jones}, {Jenkins}, {Rojo}, \&
  {Melo}}]{Jones2011}
{Jones}, M.~I., {Jenkins}, J.~S., {Rojo}, P., \& {Melo}, C.~H.~F. 2011, \aap,
  536, A71

\bibitem[{{Jones} {et~al.}(2015){Jones}, {Jenkins}, {Rojo}, {Olivares}, \&
  {Melo}}]{Jones2015}
{Jones}, M.~I., {Jenkins}, J.~S., {Rojo}, P., {Olivares}, F., \& {Melo},
  C.~H.~F. 2015, \aap, 580, A14

\bibitem[{{Kaufer} {et~al.}(1999){Kaufer}, {Stahl}, {Tubbesing},
  {N{\o}rregaard}, {Avila}, {Francois}, {Pasquini}, \& {Pizzella}}]{FEROS}
{Kaufer}, A., {Stahl}, O., {Tubbesing}, S., {et~al.} 1999, The Messenger, 95, 8

\bibitem[{{Kipping} {et~al.}(2010){Kipping}, {Bakos}, {Hartman}, {Torres},
  {Shporer}, {Latham}, {Kov{\'a}cs}, {Noyes}, {Howard}, {Fischer}, {Johnson},
  {Marcy}, {B{\'e}ky}, {Perumpilly}, {Esquerdo}, {Sasselov}, {Stefanik},
  {L{\'a}z{\'a}r}, {Papp}, \& {S{\'a}ri}}]{Kipping2010}
{Kipping}, D.~M., {Bakos}, G.~{\'A}., {Hartman}, J., {et~al.} 2010, \apj, 725,
  2017

\bibitem[{{Kunitomo} {et~al.}(2011){Kunitomo}, {Ikoma}, {Sato}, {Katsuta}, \&
  {Ida}}]{Kunitomo2011}
{Kunitomo}, M., {Ikoma}, M., {Sato}, B., {Katsuta}, Y., \& {Ida}, S. 2011,
  \apj, 737, 66

\bibitem[{{Kurucz}(1993)}]{Kurucz1993}
{Kurucz}, R. 1993, ATLAS9 Stellar Atmosphere Programs and 2 km/s grid.~Kurucz
  CD-ROM No.~13.~ Cambridge, Mass.: Smithsonian Astrophysical Observatory,
  1993., 13

\bibitem[{{Lee} {et~al.}(2013){Lee}, {Han}, \& {Park}}]{Lee2013}
{Lee}, B.-C., {Han}, I., \& {Park}, M.-G. 2013, \aap, 549, A2

\bibitem[{{Lee} {et~al.}(2014){Lee}, {Han}, {Park}, {Mkrtichian}, {Hatzes}, \&
  {Kim}}]{Lee2014}
{Lee}, B.-C., {Han}, I., {Park}, M.-G., {et~al.} 2014, \aap, 566, A67

\bibitem[{{Mayor} {et~al.}(2003){Mayor}, {Pepe}, {Queloz}, {Bouchy},
  {Rupprecht}, {Lo Curto}, {Avila}, {Benz}, {Bertaux}, {Bonfils}, {Dall},
  {Dekker}, {Delabre}, {Eckert}, {Fleury}, {Gilliotte}, {Gojak}, {Guzman},
  {Kohler}, {Lizon}, {Longinotti}, {Lovis}, {Megevand}, {Pasquini}, {Reyes},
  {Sivan}, {Sosnowska}, {Soto}, {Udry}, {van Kesteren}, {Weber}, \&
  {Weilenmann}}]{HARPS}
{Mayor}, M., {Pepe}, F., {Queloz}, D., {et~al.} 2003, The Messenger, 114, 20

\bibitem[{{M{\'e}sz{\'a}ros} {et~al.}(2012){M{\'e}sz{\'a}ros}, {Allende
  Prieto}, {Edvardsson}, {Castelli}, {Garc{\'{\i}}a P{\'e}rez}, {Gustafsson},
  {Majewski}, {Plez}, {Schiavon}, {Shetrone}, \& {de Vicente}}]{Meszaros2012}
{M{\'e}sz{\'a}ros}, S., {Allende Prieto}, C., {Edvardsson}, B., {et~al.} 2012,
  \aj, 144, 120

\bibitem[{{Mortier} {et~al.}(2013){Mortier}, {Santos}, {Sousa}, {Israelian},
  {Mayor}, \& {Udry}}]{Mortier2013}
{Mortier}, A., {Santos}, N.~C., {Sousa}, S., {et~al.} 2013, \aap, 551, A112

\bibitem[{{Mortier} {et~al.}(2014){Mortier}, {Sousa}, {Adibekyan},
  {Brand{\~a}o}, \& {Santos}}]{Mortier2014}
{Mortier}, A., {Sousa}, S.~G., {Adibekyan}, V.~Z., {Brand{\~a}o}, I.~M., \&
  {Santos}, N.~C. 2014, \aap, 572, A95

\bibitem[{{Motalebi} {et~al.}(2015){Motalebi}, {Udry}, {Gillon}, {Lovis},
  {S{\'e}gransan}, {Buchhave}, {Demory}, {Malavolta}, {Dressing}, {Sasselov},
  {Rice}, {Charbonneau}, {Collier Cameron}, {Latham}, {Molinari}, {Pepe},
  {Affer}, {Bonomo}, {Cosentino}, {Dumusque}, {Figueira}, {Fiorenzano},
  {Gettel}, {Harutyunyan}, {Haywood}, {Johnson}, {Lopez}, {Lopez-Morales},
  {Mayor}, {Micela}, {Mortier}, {Nascimbeni}, {Philips}, {Piotto}, {Pollacco},
  {Queloz}, {Sozzetti}, {Vanderburg}, \& {Watson}}]{Motalebi2015}
{Motalebi}, F., {Udry}, S., {Gillon}, M., {et~al.} 2015, \aap, 584, A72

\bibitem[{{Moutou} {et~al.}(2015){Moutou}, {Lo Curto}, {Mayor}, {Bouchy},
  {Benz}, {Lovis}, {Naef}, {Pepe}, {Queloz}, {Santos}, {S{\'e}gransan},
  {Sousa}, \& {Udry}}]{Moutou2015}
{Moutou}, C., {Lo Curto}, G., {Mayor}, M., {et~al.} 2015, \aap, 576, A48

\bibitem[{{Neves} {et~al.}(2009){Neves}, {Santos}, {Sousa}, {Correia}, \&
  {Israelian}}]{Neves2009}
{Neves}, V., {Santos}, N.~C., {Sousa}, S.~G., {Correia}, A.~C.~M., \&
  {Israelian}, G. 2009, \aap, 497, 563

\bibitem[{{Neveu-VanMalle} {et~al.}(2014){Neveu-VanMalle}, {Queloz},
  {Anderson}, {Charbonnel}, {Collier Cameron}, {Delrez}, {Gillon}, {Hellier},
  {Jehin}, {Lendl}, {Maxted}, {Pepe}, {Pollacco}, {S{\'e}gransan}, {Smalley},
  {Smith}, {Southworth}, {Triaud}, {Udry}, \& {West}}]{Neveu2014}
{Neveu-VanMalle}, M., {Queloz}, D., {Anderson}, D.~R., {et~al.} 2014, \aap,
  572, A49

\bibitem[{{Newton}(1687)}]{Newton1687}
{Newton}, I. 1687, {Philosophiae Naturalis Principia Mathematica. Auctore Js.
  Newton}

\bibitem[{{Niedzielski} {et~al.}(2016){Niedzielski}, {Villaver}, {Nowak},
  {Adam{\'o}w}, {Kowalik}, {Wolszczan}, {Deka-Szymankiewicz}, {Adamczyk}, \&
  {Maciejewski}}]{Niedzielski2016}
{Niedzielski}, A., {Villaver}, E., {Nowak}, G., {et~al.} 2016, \aap, 588, A62

\bibitem[{{Niedzielski} {et~al.}(2015{\natexlab{a}}){Niedzielski}, {Villaver},
  {Wolszczan}, {Adam{\'o}w}, {Kowalik}, {Maciejewski}, {Nowak},
  {Garc{\'{\i}}a-Hern{\'a}ndez}, {Deka}, \& {Adamczyk}}]{Niedzielski2015a}
{Niedzielski}, A., {Villaver}, E., {Wolszczan}, A., {et~al.}
  2015{\natexlab{a}}, \aap, 573, A36

\bibitem[{{Niedzielski} {et~al.}(2015{\natexlab{b}}){Niedzielski}, {Wolszczan},
  {Nowak}, {Adam{\'o}w}, {Kowalik}, {Maciejewski}, {Deka-Szymankiewicz}, \&
  {Adamczyk}}]{Niedzielski2015}
{Niedzielski}, A., {Wolszczan}, A., {Nowak}, G., {et~al.} 2015{\natexlab{b}},
  \apj, 803, 1

\bibitem[{{Nowak} {et~al.}(2013){Nowak}, {Niedzielski}, {Wolszczan},
  {Adam{\'o}w}, \& {Maciejewski}}]{Nowak2013}
{Nowak}, G., {Niedzielski}, A., {Wolszczan}, A., {Adam{\'o}w}, M., \&
  {Maciejewski}, G. 2013, \apj, 770, 53

\bibitem[{{Penev} {et~al.}(2013){Penev}, {Bakos}, {Bayliss}, {Jord{\'a}n},
  {Mohler}, {Zhou}, {Suc}, {Rabus}, {Hartman}, {Mancini}, {B{\'e}ky}, {Csubry},
  {Buchhave}, {Henning}, {Nikolov}, {Cs{\'a}k}, {Brahm}, {Espinoza}, {Conroy},
  {Noyes}, {Sasselov}, {Schmidt}, {Wright}, {Tinney}, {Addison},
  {L{\'a}z{\'a}r}, {Papp}, \& {S{\'a}ri}}]{Penev2013}
{Penev}, K., {Bakos}, G.~{\'A}., {Bayliss}, D., {et~al.} 2013, \aj, 145, 5

\bibitem[{{Pourbaix} {et~al.}(2004){Pourbaix}, {Tokovinin}, {Batten}, {Fekel},
  {Hartkopf}, {Levato}, {Morrell}, {Torres}, \& {Udry}}]{Pourbaix2004}
{Pourbaix}, D., {Tokovinin}, A.~A., {Batten}, A.~H., {et~al.} 2004, \aap, 424,
  727

\bibitem[{{Quinn} {et~al.}(2014){Quinn}, {White}, {Latham}, {Buchhave},
  {Torres}, {Stefanik}, {Berlind}, {Bieryla}, {Calkins}, {Esquerdo}, {F{\H
  u}r{\'e}sz}, {Geary}, \& {Szentgyorgyi}}]{Quinn2014}
{Quinn}, S.~N., {White}, R.~J., {Latham}, D.~W., {et~al.} 2014, \apj, 787, 27

\bibitem[{{Santos} {et~al.}(2003){Santos}, {Israelian}, {Mayor}, {Rebolo}, \&
  {Udry}}]{Santos2003}
{Santos}, N.~C., {Israelian}, G., {Mayor}, M., {Rebolo}, R., \& {Udry}, S.
  2003, \aap, 398, 363

\bibitem[{{Santos} {et~al.}(2013){Santos}, {Sousa}, {Mortier}, {Neves},
  {Adibekyan}, {Tsantaki}, {Delgado Mena}, {Bonfils}, {Israelian}, {Mayor}, \&
  {Udry}}]{Santos13}
{Santos}, N.~C., {Sousa}, S.~G., {Mortier}, A., {et~al.} 2013, \aap, 556, A150

\bibitem[{{Sato} {et~al.}(2008){Sato}, {Izumiura}, {Toyota}, {Kambe}, {Ikoma},
  {Omiya}, {Masuda}, {Takeda}, {Murata}, {Itoh}, {Ando}, {Yoshida}, {Kokubo},
  \& {Ida}}]{Sato2008}
{Sato}, B., {Izumiura}, H., {Toyota}, E., {et~al.} 2008, \pasj, 60, 539

\bibitem[{{Sato} {et~al.}(2012){Sato}, {Omiya}, {Harakawa}, {Izumiura},
  {Kambe}, {Takeda}, {Yoshida}, {Itoh}, {Ando}, {Kokubo}, \& {Ida}}]{Sato2012}
{Sato}, B., {Omiya}, M., {Harakawa}, H., {et~al.} 2012, \pasj, 64
  [\eprint[arXiv]{1207.3141}]

\bibitem[{{Sato} {et~al.}(2013{\natexlab{a}}){Sato}, {Omiya}, {Harakawa},
  {Liu}, {Izumiura}, {Kambe}, {Takeda}, {Yoshida}, {Itoh}, {Ando}, {Kokubo}, \&
  {Ida}}]{Sato2013}
{Sato}, B., {Omiya}, M., {Harakawa}, H., {et~al.} 2013{\natexlab{a}}, \pasj, 65
  [\eprint[arXiv]{1304.4328}]

\bibitem[{{Sato} {et~al.}(2013{\natexlab{b}}){Sato}, {Omiya}, {Wittenmyer},
  {Harakawa}, {Nagasawa}, {Izumiura}, {Kambe}, {Takeda}, {Yoshida}, {Itoh},
  {Ando}, {Kokubo}, \& {Ida}}]{Sato2013b}
{Sato}, B., {Omiya}, M., {Wittenmyer}, R.~A., {et~al.} 2013{\natexlab{b}},
  \apj, 762, 9

\bibitem[{{Simpson} {et~al.}(2011){Simpson}, {Faedi}, {Barros}, {Brown},
  {Collier Cameron}, {Hebb}, {Pollacco}, {Smalley}, {Todd}, {Butters},
  {H{\'e}brard}, {McCormac}, {Miller}, {Santerne}, {Street}, {Skillen},
  {Triaud}, {Anderson}, {Bento}, {Boisse}, {Bouchy}, {Enoch}, {Haswell},
  {Hellier}, {Holmes}, {Horne}, {Keenan}, {Lister}, {Maxted}, {Moulds},
  {Moutou}, {Norton}, {Parley}, {Pepe}, {Queloz}, {Segransan}, {Smith},
  {Stempels}, {Udry}, {Watson}, {West}, \& {Wheatley}}]{Simpson2011}
{Simpson}, E.~K., {Faedi}, F., {Barros}, S.~C.~C., {et~al.} 2011, \aj, 141, 8

\bibitem[{{Sneden}(1973)}]{Sneden1973}
{Sneden}, C.~A. 1973, PhD thesis, THE UNIVERSITY OF TEXAS AT AUSTIN.

\bibitem[{{Sousa} {et~al.}(2015){Sousa}, {Santos}, {Adibekyan}, {Delgado-Mena},
  \& {Israelian}}]{Sousa2015a}
{Sousa}, S.~G., {Santos}, N.~C., {Adibekyan}, V., {Delgado-Mena}, E., \&
  {Israelian}, G. 2015, \aap, 577, A67

\bibitem[{{Sousa} {et~al.}(2012){Sousa}, {Santos}, \& {Israelian}}]{Sousa2012}
{Sousa}, S.~G., {Santos}, N.~C., \& {Israelian}, G. 2012, \aap, 544, A122

\bibitem[{{Sousa} {et~al.}(2007){Sousa}, {Santos}, {Israelian}, {Mayor}, \&
  {Monteiro}}]{Sousa2007}
{Sousa}, S.~G., {Santos}, N.~C., {Israelian}, G., {Mayor}, M., \& {Monteiro},
  M.~J.~P.~F.~G. 2007, \aap, 469, 783

\bibitem[{{Sousa} {et~al.}(2011){Sousa}, {Santos}, {Israelian}, {Mayor}, \&
  {Udry}}]{Sousa2011}
{Sousa}, S.~G., {Santos}, N.~C., {Israelian}, G., {Mayor}, M., \& {Udry}, S.
  2011, \aap, 533, A141

\bibitem[{{Sousa} {et~al.}(2008){Sousa}, {Santos}, {Mayor}, {Udry},
  {Casagrande}, {Israelian}, {Pepe}, {Queloz}, \& {Monteiro}}]{Sousa2008a}
{Sousa}, S.~G., {Santos}, N.~C., {Mayor}, M., {et~al.} 2008, \aap, 487, 373

\bibitem[{{Takeda} {et~al.}(2008){Takeda}, {Sato}, \& {Murata}}]{Takeda2008}
{Takeda}, Y., {Sato}, B., \& {Murata}, D. 2008, \pasj, 60, 781

\bibitem[{{Torres} {et~al.}(2010){Torres}, {Andersen}, \&
  {Gim{\'e}nez}}]{Torres2010}
{Torres}, G., {Andersen}, J., \& {Gim{\'e}nez}, A. 2010, \aapr, 18, 67

\bibitem[{{Torres} {et~al.}(2012){Torres}, {Fischer}, {Sozzetti}, {Buchhave},
  {Winn}, {Holman}, \& {Carter}}]{Torres2012}
{Torres}, G., {Fischer}, D.~A., {Sozzetti}, A., {et~al.} 2012, \apj, 757, 161

\bibitem[{{Torres} {et~al.}(2008){Torres}, {Winn}, \& {Holman}}]{Torres2008}
{Torres}, G., {Winn}, J.~N., \& {Holman}, M.~J. 2008, \apj, 677, 1324

\bibitem[{{Tsantaki} {et~al.}(2013){Tsantaki}, {Sousa}, {Adibekyan}, {Santos},
  {Mortier}, \& {Israelian}}]{Tsantaki2013}
{Tsantaki}, M., {Sousa}, S.~G., {Adibekyan}, V.~Z., {et~al.} 2013, \aap, 555,
  A150

\bibitem[{{Valenti} \& {Fischer}(2005)}]{Valenti2005}
{Valenti}, J.~A. \& {Fischer}, D.~A. 2005, \apjs, 159, 141

\bibitem[{{Vanderburg} {et~al.}(2015){Vanderburg}, {Montet}, {Johnson},
  {Buchhave}, {Zeng}, {Pepe}, {Collier Cameron}, {Latham}, {Molinari}, {Udry},
  {Lovis}, {Matthews}, {Cameron}, {Law}, {Bowler}, {Angus}, {Baranec},
  {Bieryla}, {Boschin}, {Charbonneau}, {Cosentino}, {Dumusque}, {Figueira},
  {Guenther}, {Harutyunyan}, {Hellier}, {Kuschnig}, {Lopez-Morales}, {Mayor},
  {Micela}, {Moffat}, {Pedani}, {Phillips}, {Piotto}, {Pollacco}, {Queloz},
  {Rice}, {Riddle}, {Rowe}, {Rucinski}, {Sasselov}, {S{\'e}gransan},
  {Sozzetti}, {Szentgyorgyi}, {Watson}, \& {Weiss}}]{Vanderburg2015}
{Vanderburg}, A., {Montet}, B.~T., {Johnson}, J.~A., {et~al.} 2015, \apj, 800,
  59

\bibitem[{{West} {et~al.}(2016){West}, {Hellier}, {Almenara}, {Anderson},
  {Barros}, {Bouchy}, {Brown}, {Collier Cameron}, {Deleuil}, {Delrez}, {Doyle},
  {Faedi}, {Fumel}, {Gillon}, {G{\'o}mez Maqueo Chew}, {H{\'e}brard}, {Jehin},
  {Lendl}, {Maxted}, {Pepe}, {Pollacco}, {Queloz}, {S{\'e}gransan}, {Smalley},
  {Smith}, {Southworth}, {Triaud}, \& {Udry}}]{West2016}
{West}, R.~G., {Hellier}, C., {Almenara}, J.-M., {et~al.} 2016, \aap, 585, A126

\bibitem[{{Wilson} {et~al.}(2016){Wilson}, {H{\'e}brard}, {Santos}, {Sahlmann},
  {Montagnier}, {Astudillo-Defru}, {Boisse}, {Bouchy}, {Rey}, {Arnold},
  {Bonfils}, {Bourrier}, {Courcol}, {Deleuil}, {Delfosse}, {D{\'{\i}}az},
  {Ehrenreich}, {Forveille}, {Moutou}, {Pepe}, {Santerne}, {S{\'e}gransan}, \&
  {Udry}}]{Wilson2016}
{Wilson}, P.~A., {H{\'e}brard}, G., {Santos}, N.~C., {et~al.} 2016, \aap, 588,
  A144

\bibitem[{{Zhou} {et~al.}(2014){Zhou}, {Bayliss}, {Penev}, {Bakos}, {Hartman},
  {Jord{\'a}n}, {Mancini}, {Mohler}, {Csubry}, {Ciceri}, {Brahm}, {Rabus},
  {Buchhave}, {Henning}, {Suc}, {Espinoza}, {B{\'e}ky}, {Noyes}, {Schmidt},
  {Butler}, {Shectman}, {Thompson}, {Crane}, {Sato}, {Cs{\'a}k},
  {L{\'a}z{\'a}r}, {Papp}, {S{\'a}ri}, \& {Nikolov}}]{Zhou2014}
{Zhou}, G., {Bayliss}, D., {Penev}, K., {et~al.} 2014, \aj, 147, 144

\bibitem[{{Zieli{\'n}ski} {et~al.}(2012){Zieli{\'n}ski}, {Niedzielski},
  {Wolszczan}, {Adam{\'o}w}, \& {Nowak}}]{Zielinski2012}
{Zieli{\'n}ski}, P., {Niedzielski}, A., {Wolszczan}, A., {Adam{\'o}w}, M., \&
  {Nowak}, G. 2012, \aap, 547, A91

\end{thebibliography}

\begin{appendix}
\section{Updated parameters of 50 planet hosts}

\onecolumn
\begin{longtable}{lllrlclr}
    \caption{\label{tab:results} Derived parameters for the 50 stars in our
             sample. The S/N was measured by ARES.}\\
    \hline\hline
    Star  &  $T_\mathrm{eff}$ (K)  &  $\log g$ (dex)  &  $[\ion{Fe}/\ion{H}]$ (dex)  &  $\xi_\mathrm{micro}$ (km/s)  &  $\xi_\mathrm{micro}$ fixed?  &  Instrument  &  S/N \\
    \hline
    \endfirsthead
    \caption{continued.}\\
    \hline
    \endhead
    \hline
    \endfoot
    \object{BD -11 4672}    &   $4553 \pm 75 $   &  $4.87 \pm 0.51$                  &  $-0.30 \pm 0.02$  &  $0.14 \pm 0.07$  & yes  &  FIES             &  487  \\
    \object{BD +49  828}    &   $5015 \pm 36 $   &  $2.87 \pm 0.09$\tablefootmark{a} &  $-0.01 \pm 0.03$  &  $1.48 \pm 0.04$  & no   &  FIES             &  567  \\
    \object{GJ 785}         &   $5087 \pm 48 $   &  $4.42 \pm 0.10$                  &  $-0.01 \pm 0.03$  &  $0.69 \pm 0.10$  & no   &  HARPS            &  801  \\
    \object{HATS-1}         &   $5969 \pm 46 $   &  $4.39 \pm 0.06$                  &  $-0.04 \pm 0.04$  &  $1.06 \pm 0.08$  & no   &  UVES             &  155  \\
    \object{HATS-5}         &   $5383 \pm 91 $   &  $4.41 \pm 0.22$                  &  $ 0.08 \pm 0.06$  &  $0.91 \pm 0.14$  & no   &  UVES             &  158  \\
    \object{HAT-P-12}       &   $4642 \pm 106$   &  $4.53 \pm 0.27$                  &  $-0.26 \pm 0.06$  &  $0.28 \pm 0.63$  & no   &  FIES             &  185  \\
    \object{HAT-P-24}       &   $6470 \pm 181$   &  $4.33 \pm 0.27$                  &  $-0.41 \pm 0.10$  &  $1.40 \pm 0.03$  & yes  &  UVES             &  158  \\
    \object{HAT-P-39}       &   $6745 \pm 236$   &  $4.39 \pm 0.47$                  &  $-0.21 \pm 0.12$  &  $1.53 \pm 0.04$  & yes  &  UVES             &  127  \\
    \object{HAT-P-42}       &   $5903 \pm 66 $   &  $4.29 \pm 0.10$\tablefootmark{a} &  $ 0.34 \pm 0.05$  &  $1.19 \pm 0.08$  & no   &  UVES             &  130  \\
    \object{HAT-P-46}       &   $6421 \pm 121$   &  $4.53 \pm 0.14$\tablefootmark{a} &  $ 0.16 \pm 0.09$  &  $1.67 \pm 0.18$  & no   &  UVES             &  208  \\[5pt]
    \object{HD 120084}      &   $4969 \pm 40 $   &  $2.94 \pm 0.14$\tablefootmark{a} &  $ 0.12 \pm 0.03$  &  $1.41 \pm 0.04$  & no   &  ESPaDOnS         &  852  \\
    \object{HD 192263}      &   $4946 \pm 46 $   &  $4.61 \pm 0.14$                  &  $-0.05 \pm 0.02$  &  $0.66 \pm 0.12$  & no   &  HARPS            &  415  \\
    \object{HD 219134}      &   $4767 \pm 70 $   &  $4.57 \pm 0.17$                  &  $ 0.00 \pm 0.04$  &  $0.59 \pm 0.24$  & no   &  ESPaDOnS         &  725  \\
    \object{HD 220842}      &   $5999 \pm 39 $   &  $4.30 \pm 0.06$\tablefootmark{a} &  $-0.08 \pm 0.03$  &  $1.21 \pm 0.05$  & no   &  FIES             &  459  \\
    \object{HD 233604}      &   $4954 \pm 46 $   &  $2.86 \pm 0.11$\tablefootmark{a} &  $-0.14 \pm 0.04$  &  $1.61 \pm 0.05$  & no   &  FIES             &  314  \\
    \object{HD 283668}      &   $4841 \pm 73 $   &  $4.51 \pm 0.18$                  &  $-0.74 \pm 0.04$  &  $0.16 \pm 0.61$  & no   &  FIES             &  592  \\
    \object{HD 285507}      &   $4620 \pm 126$   &  $4.72 \pm 0.61$                  &  $ 0.04 \pm 0.06$  &  $0.74 \pm 0.43$  & no   &  UVES             &  239  \\
    \object{HD 5583}        &   $4986 \pm 35 $   &  $2.87 \pm 0.09$\tablefootmark{a} &  $-0.35 \pm 0.03$  &  $1.62 \pm 0.04$  & no   &  FIES             &  933  \\
    \object{HD 81688}       &   $4903 \pm 21 $   &  $2.70 \pm 0.05$\tablefootmark{a} &  $-0.21 \pm 0.02$  &  $1.54 \pm 0.02$  & no   & \tablefootmark{b} & 1350, 860  \\
    \object{HD 82886}       &   $5123 \pm 18 $   &  $3.30 \pm 0.04$\tablefootmark{a} &  $-0.25 \pm 0.01$  &  $1.16 \pm 0.02$  & no   & \tablefootmark{c} & 1198,1294  \\[5pt]
    \object{HD 87883}       &   $4917 \pm 68 $   &  $4.53 \pm 0.19$                  &  $ 0.02 \pm 0.03$  &  $0.46 \pm 0.21$  & no   &  ESPaDOnS         &  753  \\
    \object{HIP 107773}     &   $4957 \pm 49 $   &  $2.83 \pm 0.09$\tablefootmark{a} &  $ 0.04 \pm 0.04$  &  $1.49 \pm 0.05$  & no   &  UVES             &  218  \\
    \object{HIP 11915}      &   $5770 \pm 14 $   &  $4.33 \pm 0.03$                  &  $-0.06 \pm 0.01$  &  $0.95 \pm 0.02$  & no   &  HARPS            &  709  \\
    \object{HIP 116454}     &   $5042	\pm 72 $   &  $4.69 \pm 0.15$                  &  $-0.16 \pm 0.03$  &  $0.71 \pm 0.17$  & no   &  UVES             &  412  \\
    \object{HR 228}         &   $5042 \pm 42 $   &  $3.30 \pm 0.09$\tablefootmark{a} &  $ 0.07 \pm 0.03$  &  $1.14 \pm 0.04$  & no   &  UVES             &  400  \\
    \object{KELT-6}         &   $6246 \pm 88 $   &  $4.22 \pm 0.09$\tablefootmark{a} &  $-0.22 \pm 0.06$  &  $1.66 \pm 0.13$  & no   &  FIES             &  374  \\
    \object{Kepler-37}      &   $5378 \pm 53 $   &  $4.47 \pm 0.12$                  &  $-0.23 \pm 0.04$  &  $0.58 \pm 0.13$  & no   &  FIES             &  205  \\
    \object{Kepler-444}     &   $5111 \pm 43 $   &  $4.50 \pm 0.13$                  &  $-0.51 \pm 0.03$  &  $0.37 \pm 0.15$  & no   &  FIES             &  675  \\
    \object{mu Leo}         &   $4605 \pm 94 $   &  $2.61 \pm 0.26$\tablefootmark{a} &  $ 0.25 \pm 0.06$  &  $1.64 \pm 0.11$  & no   &  ESPaDOnS         &  354  \\[5pt]
    \object{ome Ser}        &   $4928 \pm 35 $   &  $2.69 \pm 0.06$\tablefootmark{a} &  $-0.11 \pm 0.03$  &  $1.55 \pm 0.04$  & no   &  FIES             & 1168  \\
    \object{omi UMa}        &   $5499 \pm 52 $   &  $3.36 \pm 0.07$\tablefootmark{a} &  $-0.01 \pm 0.05$  &  $1.98 \pm 0.06$  & no   &  ESPaDOnS         &  527  \\
    \object{Qatar-}2        &   $4637 \pm 316$   &  $4.53 \pm 0.62$                  &  $ 0.09 \pm 0.17$  &  $0.63 \pm 0.83$  & no   &  UVES             &   97  \\
    \object{SAND364}        &   $4457 \pm 104$   &  $2.26 \pm 0.20$\tablefootmark{a} &  $-0.04 \pm 0.06$  &  $1.60 \pm 0.11$  & no   &  UVES             &  220  \\
    \object{TYC+1422-614-1} &   $4908 \pm 41 $   &  $2.90 \pm 0.12$\tablefootmark{a} &  $-0.07 \pm 0.03$  &  $1.57 \pm 0.05$  & no   &  FIES             &  506  \\
    \object{WASP-37}        &   $5917 \pm 72 $   &  $4.25 \pm 0.15$                  &  $-0.23 \pm 0.05$  &  $0.59 \pm 0.13$  & no   &  FIES             &  232  \\
    \object{WASP-44}        &   $5612 \pm 80 $   &  $4.39 \pm 0.30$                  &  $ 0.17 \pm 0.06$  &  $1.32 \pm 0.13$  & no   &  UVES             &  125  \\
    \object{WASP-52}        &   $5197 \pm 83 $   &  $4.55 \pm 0.30$                  &  $ 0.15 \pm 0.05$  &  $1.16 \pm 0.14$  & no   &  UVES             &  125  \\
    \object{WASP-58}        &   $6039 \pm 55 $   &  $4.23 \pm 0.10$                  &  $-0.09 \pm 0.04$  &  $1.12 \pm 0.08$  & no   &  FIES             &  310  \\
    \object{WASP-61}        &   $6265 \pm 168$   &  $4.21 \pm 0.21$\tablefootmark{a} &  $-0.38 \pm 0.11$  &  $1.44 \pm 0.02$  & yes  &  UVES             &  163  \\[5pt]
    \object{WASP-72}        &   $6570 \pm 85 $   &  $4.25 \pm 0.13$                  &  $ 0.15 \pm 0.06$  &  $2.30 \pm 0.15$  & no   &  UVES             &  174  \\
    \object{WASP-73}        &   $6203 \pm 32 $   &  $4.16 \pm 0.06$\tablefootmark{a} &  $ 0.20 \pm 0.02$  &  $1.66 \pm 0.04$  & np   & \tablefootmark{d} & 193,231 \\
    \object{WASP-75}        &   $6203 \pm 46 $   &  $4.42 \pm 0.22$\tablefootmark{a} &  $ 0.24 \pm 0.03$  &  $1.45 \pm 0.06$  & no   &  UVES             &  189  \\
    \object{WASP-76}        &   $6347 \pm 52 $   &  $4.29 \pm 0.08$\tablefootmark{a} &  $ 0.36 \pm 0.04$  &  $1.73 \pm 0.06$  & no   &  UVES             &  165  \\
    \object{WASP-82}        &   $6563 \pm 55 $   &  $4.29 \pm 0.10$\tablefootmark{a} &  $ 0.18 \pm 0.04$  &  $1.93 \pm 0.08$  & no   &  UVES             &  239  \\
    \object{WASP-88}        &   $6450 \pm 61 $   &  $4.24 \pm 0.06$\tablefootmark{a} &  $ 0.03 \pm 0.04$  &  $1.79 \pm 0.09$  & no   &  UVES             &  174  \\
    \object{WASP-94 A}      &   $6259 \pm 34 $   &  $4.34 \pm 0.07$\tablefootmark{a} &  $ 0.35 \pm 0.03$  &  $1.50 \pm 0.04$  & no   &  UVES             &  356  \\
    \object{WASP-94 B}      &   $6137 \pm 21 $   &  $4.42 \pm 0.05$\tablefootmark{a} &  $ 0.33 \pm 0.02$  &  $1.29 \pm 0.03$  & no   &  UVES             &  397  \\
    \object{WASP-95}        &   $5799 \pm 31 $   &  $4.29 \pm 0.05$\tablefootmark{a} &  $ 0.22 \pm 0.03$  &  $1.18 \pm 0.04$  & no   &  UVES             &  247  \\
    \object{WASP-97}        &   $5723 \pm 52 $   &  $4.24 \pm 0.07$                  &  $ 0.31 \pm 0.04$  &  $1.03 \pm 0.08$  & no   &  UVES             &  219  \\[5pt]
    \object{WASP-99}        &   $6324 \pm 89 $   &  $4.34 \pm 0.12$                  &  $ 0.27 \pm 0.06$  &  $1.83 \pm 0.12$  & no   &  UVES             &  249  \\
    \object{WASP-100}       &   $6853 \pm 209$   &  $4.15 \pm 0.26$\tablefootmark{a} &  $-0.30 \pm 0.12$  &  $1.87 \pm 0.02$  & yes  &  UVES             &  166  \\

\end{longtable}
\tablefoot{
\tablefoottext{a}{Spectroscopic $\log g$.}\\
\tablefoottext{b}{Weighted average of ESPaDoNS and FIES results. The parameters
                  are (FIES in parantheses):
                  $T_\mathrm{eff}=4870(4934)\pm30(29)$,
                  $\log g=2.50(2.73)\pm0.14(0.05)$,
                  $[\ion{Fe}/\ion{H}]=-0.26(-0.19)\pm0.03(0.02)$, and
                  $\xi_\mathrm{micro}=1.50(1.59)\pm0.03(0.03)$.}\\
\tablefoottext{c}{Weighted average of ESPaDoNS and FIES results. The parameters
                  are (FIES in parantheses):
                  $T_\mathrm{eff}=5124(5121)\pm22(29)$,
                  $\log g=3.30(3.31)\pm0.05(0.07)$,
                  $[\ion{Fe}/\ion{H}]=-0.25(-0.24)\pm0.02(0.02)$, and
                  $\xi_\mathrm{micro}=1.15(1.17)\pm0.03(0.04)$.
}\\
\tablefoottext{d}{Weighted average of UVES and FEROS results. The parameters
                  are (FEROS in parantheses):
                  $T_\mathrm{eff}=6313(6162)\pm61(37)$,
                  $\log g=4.26(4.14)\pm0.15(0.06)$,
                  $[\ion{Fe}/\ion{H}]=0.22(0.19)\pm0.04(0.03)$, and
                  $\xi_\mathrm{micro}=1.85(1.61)\pm0.08(0.04)$.
}}

\begin{longtable}{lllrll}
    \caption{\label{tab:oldSC} Previous parameters from SWEET-Cat.}\\
    \hline\hline
    Star  &  $T_\mathrm{eff}$ (K)  &  $\log g$ (dex)  &  $[\ion{Fe}/\ion{H}]$ (dex)  &  $\xi_\mathrm{micro}$ (km/s)  &  Reference \\
    \hline
    \endfirsthead
    \caption{continued.}\\
    \hline
    \endhead
    \hline
    \endfoot
    \object{BD-114672}       &    $4475 \pm 100$   &    $4.10 \pm 0.36$   &    $-0.48 \pm 0.05$   &    $0.67 \pm 0.16$   &    \citet{Moutou2015}       \\
    \object{BD +49 828}      &    $4943 \pm  30$   &    $2.85 \pm 0.09$   &    $-0.19 \pm 0.06$   &          ...         &    \citet{Niedzielski2015}  \\
    \object{GJ 785}          &    $5144 \pm  50$   &    $4.60 \pm 0.06$   &    $ 0.08 \pm 0.03$   &          ...         &    \citet{Howard2011}       \\
    \object{HATS-1}          &    $5780 \pm 100$   &    $4.40 \pm 0.08$   &    $-0.06 \pm 0.12$   &          ...         &    \citet{Penev2013}        \\
    \object{HATS-5}          &    $5304 \pm  50$   &    $4.53 \pm 0.02$   &    $ 0.19 \pm 0.08$   &          ...         &    \citet{Zhou2014}         \\
    \object{HAT-P-12}        &    $4650 \pm  60$   &    $4.61 \pm 0.02$   &    $-0.29 \pm 0.05$   &          ...         &    \citet{Lee2014}          \\
    \object{HAT-P-24}        &    $6373 \pm  80$   &    $4.29 \pm 0.04$   &    $-0.16 \pm 0.08$   &          ...         &    \citet{Kipping2010}      \\
    \object{HAT-P-39}        &    $6340 \pm 100$   &    $4.16 \pm 0.03$   &    $ 0.19 \pm 0.10$   &          ...         &    \citet{Hartman2012}      \\
    \object{HAT-P-46}        &    $6120 \pm 100$   &    $4.25 \pm 0.11$   &    $ 0.30 \pm 0.10$   &    $0.85 \pm  ...$   &    \citet{Hartman2014}      \\
    \object{HAT-P-42}        &    $5743 \pm  50$   &    $4.14 \pm 0.07$   &    $ 0.27 \pm 0.08$   &          ...         &    \citet{Boisse2013}       \\
    \object{HD 120084}       &    $4892 \pm  22$   &    $2.71 \pm 0.08$   &    $ 0.09 \pm 0.05$   &    $1.31 \pm 0.10$   &    \citet{Sato2013}         \\
    \object{HD 192263}       &    $4906 \pm  57$   &    $4.36 \pm 0.17$   &    $-0.07 \pm 0.02$   &    $0.78 \pm 0.12$   &    \citet{Tsantaki2013}     \\
    \object{HD 219134}       &    $4699 \pm  16$   &    $4.63 \pm 0.10$   &    $ 0.11 \pm 0.04$   &    $0.35 \pm 0.19$   &    \citet{Motalebi2015}     \\
    \object{HD 220074}       &    $3935 \pm 110$   &    $1.30 \pm 0.50$   &    $-0.25 \pm 0.25$   &    $1.60 \pm 0.30$   &    \citet{Lee2013}          \\
    \object{HD 220842}       &    $5920 \pm  20$   &    $4.24 \pm 0.02$   &    $-0.17 \pm 0.02$   &          ...         &    \citet{Hebrard2016}      \\
    \object{HD 233604}       &    $4791 \pm  45$   &    $2.55 \pm 0.18$   &    $-0.36 \pm 0.04$   &          ...         &    \citet{Nowak2013}        \\
    \object{HD 283668}       &    $4845 \pm  66$   &    $4.35 \pm 0.12$   &    $-0.75 \pm 0.12$   &    $0.02 \pm 0.30$   &    \citet{Wilson2016}       \\
    \object{HD 285507}       &    $4503 \pm  73$   &    $4.67 \pm 0.06$   &    $ 0.13 \pm 0.01$   &          ...         &    \citet{Quinn2014}        \\
    \object{HD 5583}         &    $4830 \pm  45$   &    $2.53 \pm 0.14$   &    $-0.50 \pm 0.18$   &          ...         &    \citet{Niedzielski2016}  \\
    \object{HD 81688}        &    $4753 \pm  15$   &    $2.22 \pm 0.05$   &    $-0.36 \pm 0.02$   &    $1.43 \pm 0.05$   &    \citet{Sato2008}         \\
    \object{HD 82886}        &    $5112 \pm  44$   &    $3.40 \pm 0.06$   &    $-0.31 \pm 0.03$   &          ...         &    \citet{Johnson2011}      \\
    \object{HD 87883}        &    $4958 \pm  44$   &    $4.56 \pm 0.06$   &    $ 0.07 \pm 0.03$   &          ...         &    \citet{Valenti2005}      \\
    \object{HIP 107773}      &    $4945 \pm 100$   &    $2.60 \pm 0.20$   &    $ 0.03 \pm 0.10$   &          ...         &    \citet{Jones2015}        \\
    \object{HIP 11915}       &    $5760 \pm   4$   &    $4.46 \pm 0.01$   &    $-0.06 \pm 0.00$   &          ...         &    \citet{Bedell2015}       \\
    \object{HIP 116454}      &    $5089 \pm  50$   &    $4.59 \pm 0.03$   &    $-0.16 \pm 0.08$   &          ...         &    \citet{Vanderburg2015}   \\
    \object{HR 228}          &    $4959 \pm  25$   &    $3.16 \pm 0.08$   &    $ 0.01 \pm 0.04$   &    $1.12 \pm 0.07$   &    \citet{Sato2013b}        \\
    \object{KELT-6}          &    $6102 \pm  43$   &    $4.07 \pm 0.06$   &    $-0.28 \pm 0.04$   &          ...         &    \citet{Collins2014}      \\
    \object{Kepler-37}       &    $5417 \pm  70$   &    $4.57 \pm 0.01$   &    $-0.32 \pm 0.07$   &          ...         &    \citet{Barclay2013}      \\
    \object{Kepler-444}      &    $5046 \pm  74$   &    $4.60 \pm 0.06$   &    $-0.55 \pm 0.07$   &          ...         &    \citet{Campante2015}     \\
    \object{mu Leo}          &    $4538 \pm  27$   &    $2.40 \pm 0.10$   &    $ 0.36 \pm 0.05$   &    $1.40 \pm 0.10$   &    \citet{Lee2014}          \\
    \object{ome Ser}         &    $4770 \pm  10$   &    $2.32 \pm 0.04$   &    $-0.24 \pm 0.02$   &    $1.34 \pm 0.04$   &    \citet{Sato2013}         \\
    \object{omi UMa}         &    $5242 \pm  10$   &    $2.64 \pm 0.03$   &    $-0.09 \pm 0.02$   &    $1.51 \pm 0.07$   &    \citet{Sato2012}         \\
    \object{Qatar-2}         &    $4645 \pm  50$   &    $4.60 \pm 0.02$   &    $-0.02 \pm 0.08$   &          ...         &    \citet{Bryan2012}        \\
    \object{SAND364}         &    $4284 \pm   9$   &    $2.20 \pm 0.06$   &    $-0.02 \pm 0.04$   &          ...         &    \citet{Brucalassi2014}   \\
    \object{TYC+1422-614-1}  &    $4806 \pm  45$   &    $2.85 \pm 0.18$   &    $-0.20 \pm 0.08$   &          ...         &    \citet{Niedzielski2015a} \\
    \object{WASP-37}         &    $5940 \pm  55$   &    $4.39 \pm 0.02$   &    $-0.40 \pm 0.12$   &          ...         &    \citet{Simpson2011}      \\
    \object{WASP-44}         &    $5400 \pm 150$   &    $4.48 \pm 0.07$   &    $ 0.06 \pm 0.10$   &          ...         &    \citet{Anderson2012}     \\
    \object{WASP-52}         &    $5000 \pm 100$   &    $4.58 \pm 0.01$   &    $ 0.03 \pm 0.12$   &          ...         &    \citet{Hebrard2013}      \\
    \object{WASP-58}         &    $5800 \pm 150$   &    $4.27 \pm 0.09$   &    $-0.45 \pm 0.09$   &          ...         &    \citet{Hebrard2013}      \\
    \object{WASP-61}         &    $6250 \pm 150$   &    $4.26 \pm 0.01$   &    $-0.10 \pm 0.12$   &          ...         &    \citet{Hellier2012}      \\
    \object{WASP-72}         &    $6250 \pm 100$   &    $4.08 \pm 0.13$   &    $-0.06 \pm 0.09$   &    $1.60 \pm 0.10$   &    \citet{Gillon2013}       \\
    \object{WASP-73}         &    $6030 \pm 120$   &    $3.92 \pm 0.08$   &    $ 0.14 \pm 0.14$   &    $1.10 \pm 0.20$   &    \citet{Delrez2014}       \\
    \object{WASP-75}         &    $6100 \pm 100$   &    $4.50 \pm 0.10$   &    $ 0.07 \pm 0.09$   &    $1.30 \pm 0.10$   &    \citet{Gomez2013}        \\
    \object{WASP-76}         &    $6250 \pm 100$   &    $4.13 \pm 0.02$   &    $ 0.23 \pm 0.10$   &    $1.40 \pm 0.10$   &    \citet{West2016}         \\
    \object{WASP-82}         &    $6490 \pm 100$   &    $3.97 \pm 0.02$   &    $ 0.12 \pm 0.11$   &    $1.50 \pm 0.10$   &    \citet{West2016}         \\
    \object{WASP-88}         &    $6430 \pm 130$   &    $4.03 \pm 0.09$   &    $-0.08 \pm 0.12$   &    $1.40 \pm 0.10$   &    \citet{Delrez2014}       \\
    \object{WASP-94 A}       &    $6170 \pm  80$   &    $4.27 \pm 0.07$   &    $ 0.26 \pm 0.15$   &          ...         &   \citet{Neveu2014}         \\
    \object{WASP-94 B}       &    $6040 \pm  90$   &    $4.26 \pm 0.06$   &    $ 0.23 \pm 0.14$   &          ...         &   \citet{Neveu2014}         \\
    \object{WASP-95}         &    $5630 \pm 130$   &    $4.38 \pm 0.03$   &    $ 0.14 \pm 0.16$   &          ...         &    \citet{Hellier2014}      \\
    \object{WASP-97}         &    $5640 \pm 100$   &    $4.43 \pm 0.03$   &    $ 0.23 \pm 0.11$   &          ...         &    \citet{Hellier2014}      \\
    \object{WASP-99}         &    $6180 \pm 100$   &    $4.12 \pm 0.03$   &    $ 0.21 \pm 0.15$   &          ...         &    \citet{Hellier2014}      \\
    \object{WASP-100}        &    $6900 \pm 120$   &    $4.04 \pm 0.11$   &    $-0.03 \pm 0.10$   &          ...         &    \citet{Hellier2014}      \\
\end{longtable}

\end{appendix}

\end{document}